\documentclass[aps]{revtex4}
\usepackage{amsfonts}
\usepackage{amsmath}
\usepackage{amssymb,epsf}

\begin{document}

\title{Expansion of magnetic neutron stars in an energy (in)dependent
spacetime}
\author{B. Eslam Panah$^{1,2,3}$\footnote{
email address: behzad.eslampanah@gmail.com}, G. H
Bordbar$^{1,3}$\footnote{ email address: ghbordbar@shirazu.ac.ir},
S. H. Hendi$^{1,3}$\footnote{ email address: hendi@shirazu.ac.ir},
R. Ruffini$^{2,4}$\footnote{ email address: ruffini@icra.it}, Z.
Rezaei$^{1}$\footnote{ email address: zrezaei@shirazu.ac.ir}, R.
Moradi$^{2,4}$\footnote{ email address: rahim.moradi@icranet.org}}
\affiliation{$^1$ Physics Department and Biruni Observatory,
College of Sciences, Shiraz
University, Shiraz 71454, Iran\\
$^2$ ICRANet, Piazza della Repubblica 10, I-65122 Pescara, Italy\\
$^3$ Research Institute for Astronomy and Astrophysics of Maragha (RIAAM),
P.O. Box 55134-441, Maragha, Iran\\
$^4$ ICRA and Dipartimento di Fisica, Sapienza Universita di Roma and ICRA,
Piazzale Aldo Moro 5, I-00185 Roma, Italy}

\begin{abstract}
Regarding the strong magnetic field of neutron stars and high energy regime
scenario which is based on high curvature region near the compact objects,
one is motivated to study magnetic neutron stars in an energy dependent
spacetime. In this paper, we show that such strong magnetic field and energy
dependency of spacetime have considerable effects on the properties of
neutron stars. We examine the variations of maximum mass and related radius,
Schwarzschild radius, average density, gravitational redshift, Kretschmann
scalar and Buchdahl theorem due to magnetic field and also energy dependency
of metric. First, it will be shown that the maximum mass and radius of
neutron stars are increasing function of magnetic field while average
density, redshift, the strength of gravity and Kretschmann scalar are
decreasing functions of it. These results are due to a repulsive-like force
behavior for the magnetic field. Next, the effects of the gravity's rainbow
will be studied and it will be shown that by increasing the rainbow
function, the neutron stars could enjoy an expansion in their structures.
Then, we obtain a new relation for the upper mass limit of a static
spherical neutron star with uniform density in gravity's rainbow (Buchdahl
limit) in which such upper limit is modified as $M_{eff}<\frac{4c^{2}R}{9G}$%
. In addition, stability and energy conditions for the equation of state of
neutron star matter are also investigated and a comparison with empirical
results is done. It is notable that the numerical study in this paper is
conducted by using the lowest order constrained variational (LOCV) approach
in the presence of magnetic field employing AV18 potential.
\end{abstract}

\maketitle

\section{Introduction}

Neutron stars with high-density baryonic matter and in the presence strong
magnetic fields are of considerable interests in astrophysics. Interior
strong magnetic field of a neutron star can form from the compression of
constant magnetic flux. This magnetic field can be as large as $10^{19}\ G$
in the core of high density inhomogeneous gravitationally bound neutron
stars \cite{Ferrer}. By comparing with the observational data, the magnetic
field strength of neutron stars is obtained about $10^{19}\ G$ \cite{Yuan}.
Besides, in the quark core of the high density neutron stars, the maximum
field may reach up to about $10^{20}\ G$ \cite{Ferrer,Tatsumi}. On the other
hand, in order to study the compact objects such as neutron and quark stars,
one has to consider the effects of their curvature in the context of general
relativity (GR). The first hydrostatic equilibrium equation (HEE) for stars
was obtained by Tolman, Oppenheimer, and Volkoff (TOV) \cite%
{TolmanI,TolmanII,Oppenheimer}. Employing TOV equation for investigating the
properties of compact objects has been done by many authors \cite%
{Silbar,Narain,BordbarBY,BoonsermVW,LiX,Oliveira,He,Yunes}.

Strong magnetic fields have important influences on the structure of compact
objects \cite%
{Heyl,Cardall,Kiuchi,Ciolfi,Lopes,Chirenti,Cheoun,Belvedere,Neilsen,LopesDM}%
. Using magnetohydrodynamic simulation in full GR, the stability of neutron
stars with toroidal magnetic fields has been studied \cite{Kiuchi}. It is
found that non-rotating neutron stars are dynamically unstable for some
value of magnetic field strength. Considering the equation of state (EoS) of
the nuclear matter within a relativistic model subjected to a strong
magnetic field, it has been shown that, the influence of magnetic field can
increase the mass by more than $10\%$ for stars with hyperons interior \cite%
{Lopes}. Employing an effective field theory model for the nuclear EoS and
by considering two magnetic field geometries (tangled field and a force-free
field), Kamiab et al \cite{Kamiab} have shown that these fields have
specific effects on the mass of neutron stars, so that it leads to
increasing of the maximum mass more that $30\%$. The effects of a strong
magnetic field on the nuclear EoS has been studied in \cite{Broderick}.
Cardall et al. \cite{Cardall} have investigated static neutron stars by
considering polodal magnetic fields with a simple class of electric current
distribution. They have used the nuclear EoS evaluated by Broderick et al.
in Ref. \cite{Broderick}, and found that the magnetic field increases the
maximum mass of neutron stars, Astashenok et al. \cite{Astashenok} studied
neutron stars in the context of F(R) gravity. They have used a nuclear EoS
which was similar to the one extracted by Broderick et al. \cite{Broderick},
and by assuming the global effect of magnetic field pressure, they found
that the magnetic field can also increase the maximum mass of neutron star.
The effects of strong magnetic fields on the structure of neutron stars in a
perturbative $F(R)$\ gravity and also a quantum hydrodynamics model have
been investigated in \cite{Cheoun}. It is shown that there exists a set of
parameters for the modified gravity and magnetic field strength in which
even a soft EoS can accommodate a large maximum neutron star mass through
the modified mass-radius relation. In the framework of GR and relativistic
mean field theory, employing the coupled Einstein-Maxwell-Thomas-Fermi
equations, a model for both static and uniformly rotating neutron stars has
been developed \cite{Belvedere}. It is believed that the use of realistic
parameters of rotating neutron stars obtained from numerical integration of
the self-consistent axisymmetric general relativistic equations of
equilibrium leads to the values of magnetic field and radiation efficiency
of pulsars very different from estimates based on fiducial parameters.

In solar system scale or other moderate gravitational regimes, GR is an
effective theory with accurate results, whereas in context of strong gravity
(near the compact objects), this theory has problems various in which we
address some of them. From the cosmological point of view, the usual GR
cannot explain the accelerating expansion of the universe \cite%
{PerlmutterI,PerlmutterII,Riess}.\ On the other hand, in order to construct
a quantum theory of gravity, we have to evaluate the validity of theory in
the ultraviolet (UV) regime. It was shown that GR is valid in infrared (IR)
limit while in UV regime, it needs to improve for producing accurate
results. Therefore, in order to include the UV regime, we have to modify GR.
Horava-Lifshitz gravity \cite{Horava} and gravity's rainbow \cite{MagueijoII}
are two theories which include the UV regime by considering a modification
of the usual energy-momentum relation. Among these two theories, gravity's
rainbow has more interesting results. Garattini and Saridakis\ \cite%
{GarattiniSar} have shown that considering a suitable choice of the rainbow
functions, the Horava-Lifshitz gravity can be related to the gravity's
rainbow.

Gravity's rainbow can be originate from the generalization of double special
relativity to the curved spacetime. The doubly special relativity \cite%
{Amelino-Camelia} is the special relativity when we consider another upper
limit constant to this theory. Indeed, in doubly special relativity there
are two fundamental constants; the velocity of light ($c$), and the Planck
energy ($E_{p}$). In doubly special relativity, a particle can not attain
velocity and energy larger than $c$\ and $E_{p}$, respectively ($v<c$\ and $%
E<E_{p}$, in which $v$ and $E$\ are, respectively, the velocity and the
energy of a test particle). In this theory, the metric specifying spacetime
is constructed by employing the spectrum of energy that particles can obtain
\cite{Peng}. Generalization of doubly special relativity in the presence of
gravity leads to double GR or gravity's rainbow \cite{MagueijoII}. It is
notable that, the gravity's rainbow reduces to GR in the IR limit, in which
the energy of particles has no direct effect on the curvature.

\textbf{\ }It was pointed out that although gravity's rainbow modifies
thermodynamical aspects of the black objects \cite{Ali,HendiFEP,HendiPEM},
the uncertainty principle is valid \cite{LingLZ,LiH}. The effects of this
gravity on information paradox posed by Hawking radiation and black holes
evaporation have been studied in Ref. \cite{AliFM}. Also, this theory has
been investigated in context of black holes with different configurations
\cite{Galan,Gim}. Another interesting result of this gravity is related to
early universe. Indeed by considering the energy dependent
Friedmann-Robertson-Walker metric, one can study the early universe without
big bang singularity \cite%
{NonsingularI,NonsingularII,NonsingularIII,NonsingularIV}. Therefore, it is
important to investigate the structure of other compact objects such as
neutron and quark stars in this theory.

In the present work, firstly, we want to study the effects of strong
magnetic field on the structure of neutron star in Einstein gravity. Then,
by generalizing Einstein gravity to gravity's rainbow in the present of
strong magnetic fields, we will simultaneously investigated the effects of
these two factors (magnetic field and rainbow function) on the structure of
neutron star matter.

\section{Structure properties of magnetized neutron stars}

\label{STEN}

In this paper, we are going to investigate the properties of the magnetized
neutron stars using the TOV equation and the EoS introduced in Appendix (see
Fig. \ref{Fig1a} in Appendix B). The TOV equation in Einstein gravity is
given by \cite{TolmanI,TolmanII,Oppenheimer}
\begin{equation}
\frac{dP}{dr}=\frac{G\left( c^{2}M+4\pi r^{3}P\right) }{c^{2}r\left(
2GM-c^{2}r\right) }\left( c^{2}\rho +P\right) ,  \label{TOVEN}
\end{equation}%
where $M=\int 4\pi r^{2}\rho (r)dr$, also $P$ and $\rho $ are, respectively,
the pressure and density of the fluid which are measured by local observer. $%
G$ in the above equation is the gravitational constant.

\begin{figure*}[tbp]
$%
\begin{array}{cc}
\epsfxsize=8cm \epsffile{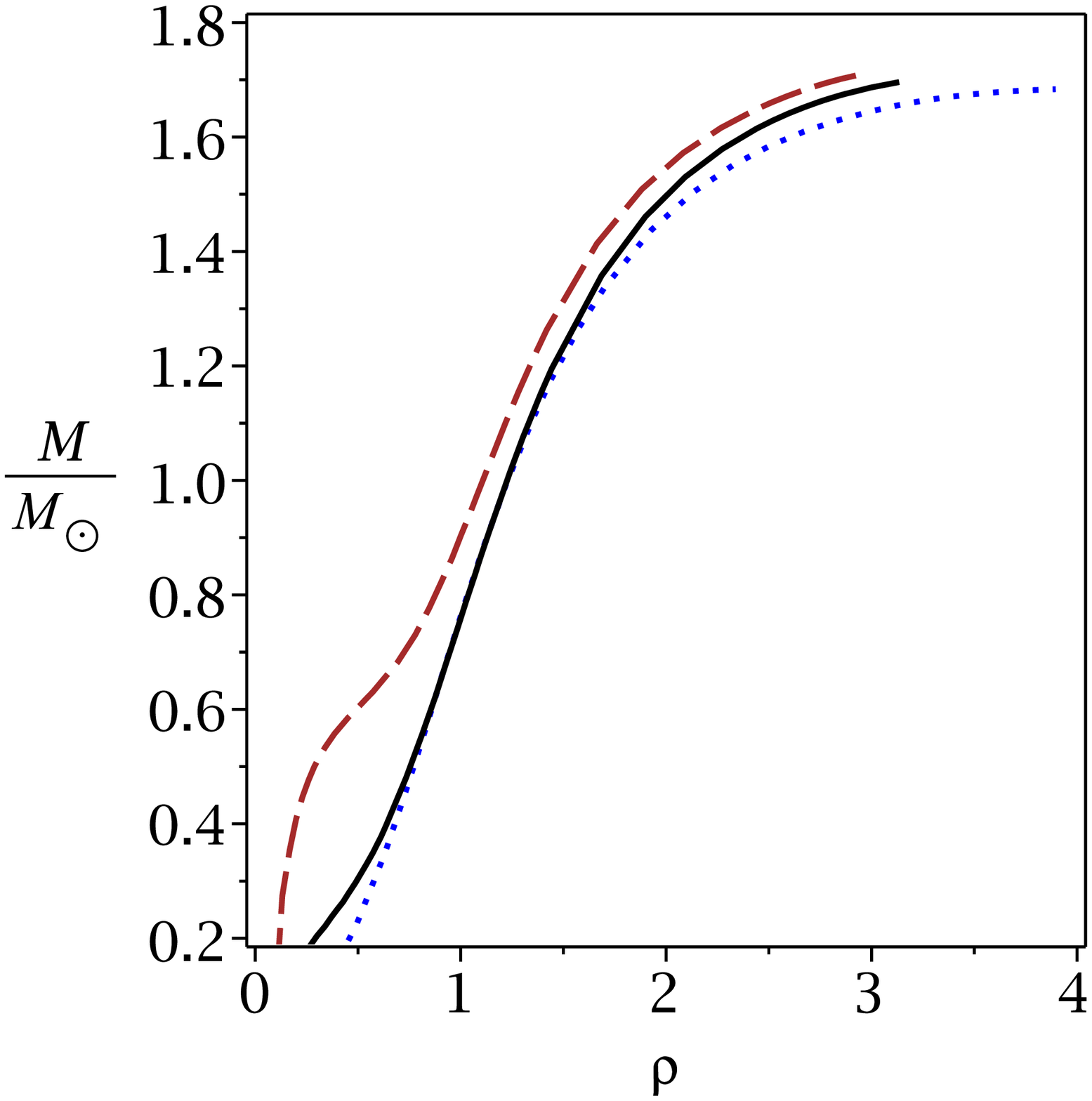} & \epsfxsize=8cm %
\epsffile{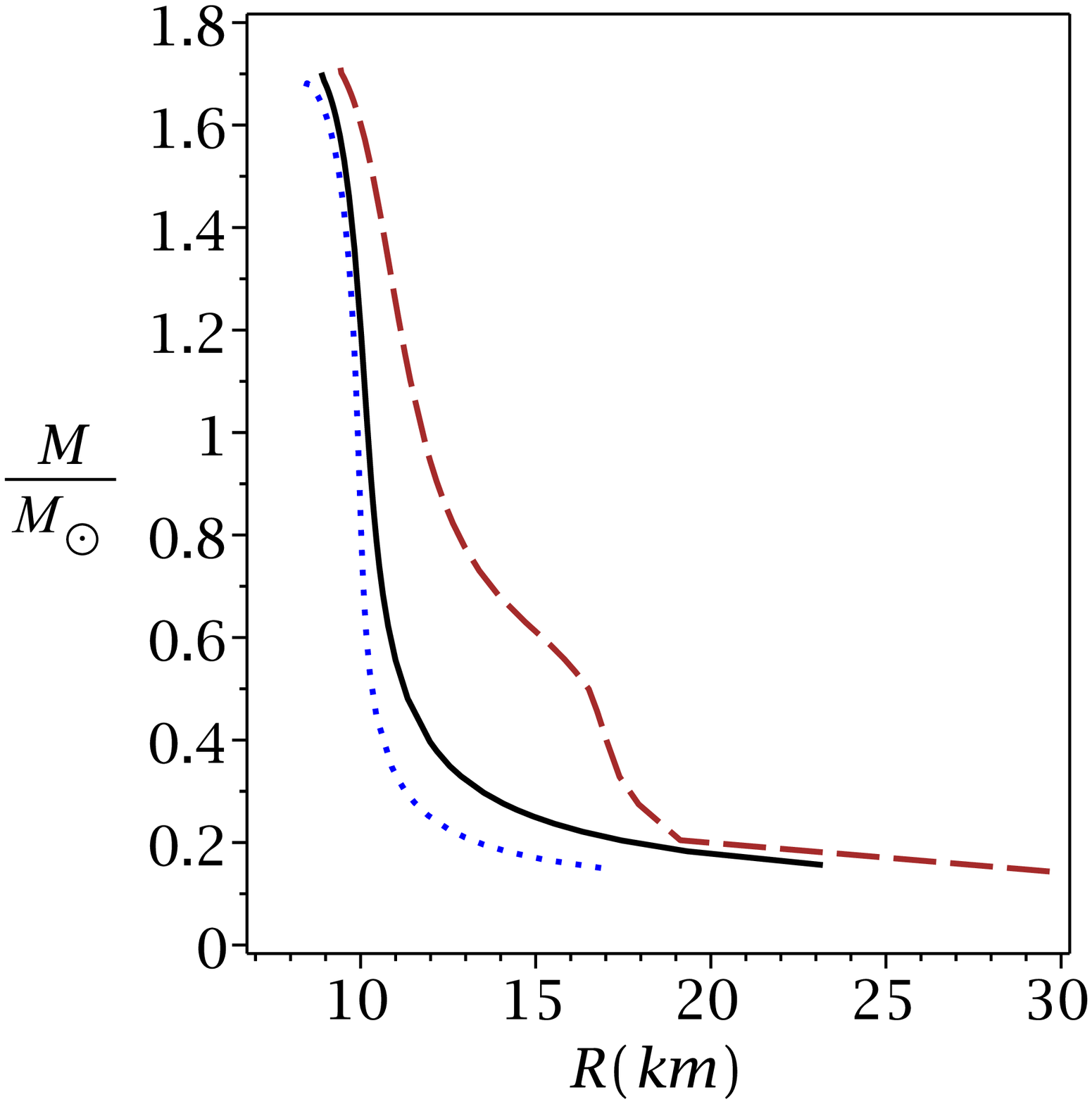}%
\end{array}
$%
\caption{Gravitational mass versus the density $(10^{15}g$ $cm^{-3})$ (right
diagram) and radius (left diagram), for $B=0$ (doted line), $B=5\times
10^{18}G$ (continuous line), and $B=1\times 10^{19}G$ (dashed line).}
\label{Fig1}
\end{figure*}

Selecting central density under the boundary conditions $P(0)=P_{c}$, $%
m(0)=0 $, we integrate the TOV equation outwards to radius $r=R$, at which $%
P $ vanishes. This yields the radius $R$ and mass $M=m(R)$ (see Ref.~\cite%
{Shapiro}). The effects of magnetic field on the gravitational mass and
radius of neutron star are presented in table \ref{tab2}. Also, the effects
of magnetic field on the diagrams related to mass-central mass density ($%
M-\rho $) and mass-radius ($M-R$) relations are presented in Fig. \ref{Fig1}%
. We found that the maximum mass and radius of magnetized neutron stars
increase when the magnetic field grows. In order to more investigation the
effect of magnetic field on neutron star, we study other interesting
properties of our solution.

One of interesting quantities of neutron star is related to the average
density of the magnetized neutron star can be written as%
\begin{equation}
\overline{\rho }=\frac{3M}{4\pi R^{3}},
\end{equation}%
the obtained results of average density for different magnetic fields are
interesting. According to table \ref{tab2} one finds that the average
density decreases when magnetic field increases. In other words, growing
magnetic field, unlike gravitational attraction, leads to more expansions of
neutron stars.

Another important quantity which is related to the compactness of a
spherical object. It can be defined as the ratio of the Schwarzschild radius
to the radius of object,
\begin{equation}
\sigma =\frac{R_{Sch}}{R},  \label{sigma}
\end{equation}%
which may be interpreted as the strength of gravity. Also the Schwarzschild
radius is $R_{Sch}=\frac{2GM}{c^{2}}$. Another quantity which shows the
strength of gravity is related to the spacetime curvature. In Schwarzschild
metric, the components of Ricci tensor ($R_{\mu \nu }$) and the Ricci scalar
($R$) are zero outside the star and these quantities do not give us any
information about the spacetime curvature. Therefore, we use another
quantity in order to more investigation of the curvature of spacetime. The
quantity which can help us to understand the curvature of spacetime is the
Riemann tensor. According to this fact the Riemann tensor may have more
components\ and for simplicity we can study the Kretschmann scalar for
measuring of the curvature in vacuum. Therefore, the curvature at the
surface of a neutron star in Einstein gravity is given as \cite{Psaltis,Eksi}
\begin{equation}
K=\sqrt{R_{\mu \nu \alpha \beta }R^{\mu \nu \alpha \beta }}=\frac{4\sqrt{3}GM%
}{c^{2}R^{3}},  \label{K}
\end{equation}

The numerical results confirm that by increasing the magnetic field, the
strength of gravity decreases and this may increase the radius of these
stars.

Another known parameter of neutron stars is the gravitational redshift which
can be written in the following form
\begin{equation}
z=\frac{1}{\sqrt{1-\frac{2GM}{c^{2}R}}}-1,
\end{equation}%
as one can see, increasing $B$, leads to decreasing the redshift.

\begin{figure}[tbp]
\epsfxsize=12cm \centerline{\epsffile{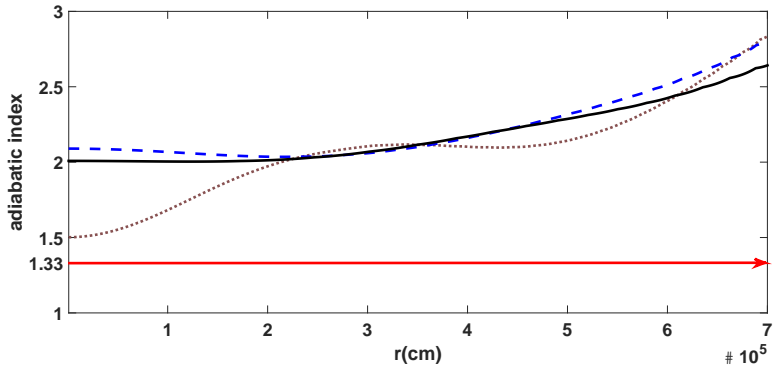}}
\caption{The adiabatic index versus the radius for $B=0$ (doted line), $%
B=5\times 10^{18}G$ (dashed line), and $B=1\times 10^{19}G $ (continuous
line).}
\label{Fig3}
\end{figure}

The dynamical stability of the stellar model against the infinitesimal
radial adiabatic perturbation was introduced by Chandrasekhar \cite%
{Chandrasekhar}. This stability condition was developed and applied to
astrophysical cases by many authors \cite{BardeenTM,Kuntsem,Mak,Kalam}. The
dynamical stability condition is satisfied when the adiabatic index, $\gamma
$, is larger than $\frac{4}{3}$, i.e, $\gamma >\frac{4}{3}=1.33$, everywhere
within the isotropic star. The adiabatic index is defined in the following
form
\begin{equation}
\gamma =\frac{\rho c^{2}+P}{c^{2}P}\frac{dP}{d\rho }.
\end{equation}%
In order to investigate the dynamical stability of magnetized neutron star,
we plot the adiabatic index versus the radius in Fig. \ref{Fig3}. As one can
see, these stars enjoy interior dynamical stability.
\begin{table*}[tbp]
\caption{Properties of neutron star with different magnetic fields.}
\label{tab2}
\begin{center}
\begin{tabular}{ccccccccc}
\hline\hline
$B(G)$ & ${M_{max}}\ (M_{\odot})$ & $R\ (km)$ & $R_{Sch}\ (km)$ & $\overline{%
\rho }$ $(10^{15}g$ $cm^{-3})$ & $\sigma (\frac{R_{Sch}}{R})$ & $z$ & $%
K(10^{-7}$ $m^{-2})$ & $\frac{4c^{2}R}{9G}\ (M_{\odot})$ \\ \hline\hline
$0$ & $1.68$ & $8.42$ & $4.95$ & $1.34$ & $0.59$ & $0.56$ & $0.29$ & $2.54$
\\ \hline
$5\times 10^{18}$ & $1.70$ & $8.88$ & $5.01$ & $1.15$ & $0.56$ & $0.51$ & $%
0.25$ & $2.68$ \\ \hline
$1\times 10^{19}$ & $1.71$ & $9.36$ & $5.04$ & $0.99$ & $0.54$ & $0.47$ & $%
0.21$ & $2.82$ \\ \hline\hline
&  &  &  &  &  &  &  &
\end{tabular}%
\end{center}
\end{table*}

Here, we want to investigate the upper mass limit of a static spherical
neutron star with uniform density in GR, so-called Buchdahl theorem \cite%
{Buchdahl}. The GR compactness limit is given by \cite{Buchdahl}
\begin{equation}
M\leq \frac{4c^{2}R}{9G},  \label{bach}
\end{equation}%
in which the upper mass limit is $M_{\max }=\frac{4c^{2}R}{9G}$. Our
numerical results confirm that the obtained mass of magnetic neutron stars
in GR respects Eq. (\ref{bach}).

Studying Fig. \ref{Fig1a} (Appendix B), shows that the core of neutron star
is more rigid when the magnetic field increases, whiles the radius of
magnetic neutron star is an increasing function of magnetic field, but the
average density, compactness, redshift and Kretschmann scalar are decreasing
functions of it. Therefore, by increasing the magnetic field, although core
of the neutron star remains compact, the outer layers expand. As a results,
the magnetic field has repulsive property.

Theoretically, it has been shown that the magnetic field contributes
directly to the pressure in the Einstein equations in which pressure fits
with square magnetic field ($P_{B}\propto B^{2}$) , see Refs. \cite%
{LopesDM,Astashenok,Moradi:2017alp,Ray:2003gt} for more details. This shows
that when magnetic field is increased, associated magnetic pressure, due to
its repulsive nature, increases the radius of neutron star.\ In other words,
by adding the magnetic field to our system, its pressure increases. Such
increasing in the pressure leads to increasing the repulsive force and as a
result the system will be unstable. In order to have a stable case, the mass
of the neutron star must increase. Therefore, it is a direct effect that
when magnetic field is increased, the total mass of system is increased, too.

In the next section, we want to evaluate the effects of gravity's rainbow on
the structure of neutron stars.

\section{Structure properties of neutron stars in gravity's rainbow}

\label{STRainbow}

Here we introduce the spherical symmetric metric in gravity's rainbow as%
\begin{equation}
ds^{2}=-\frac{\Psi (r)}{F^{2}(\varepsilon )}dt^{2}+\frac{dr^{2}}{%
H^{2}(\varepsilon )\Psi (r)}+r^{2}\left( d\theta ^{2}+\sin ^{2}(\theta
)d\varphi ^{2}\right) ,  \label{metric}
\end{equation}%
in which $H(\varepsilon )$\ and $F(\varepsilon )$\ are rainbow functions. In
order to study the structure of neutron stars in gravity's rainbow, firstly
we should obtain the HEE in this gravity. The HEE in gravity's rainbow by
considering the metric (\ref{metric}) and the following field equation
\begin{equation}
G_{\mu }^{\nu }=\frac{8\pi G}{c^{4}}T_{\mu }^{\nu },  \label{fieldEq}
\end{equation}%
where $G_{\mu }^{\nu }$\ and $T_{\mu }^{\nu }$\ are the Einstein and
energy-momentum tensors, respectively, is given by \cite{HendiBEP}
\begin{equation}
\frac{dP}{dr}=\frac{G\left( c^{2}M_{eff}H^{2}(\varepsilon )+4\pi
r^{3}P\right) }{c^{2}r\left( 2GM_{eff}-c^{2}r\right) H^{2}(\varepsilon )}%
\left( c^{2}\rho +P\right) ,  \label{TOVT}
\end{equation}%
where%
\begin{equation}
M_{eff}\left( r,\varepsilon \right) =\int \frac{4\pi }{H^{2}(\varepsilon )}%
r^{2}\rho (r)dr.  \label{dMass}
\end{equation}

Here, we consider an upper limit for energies which $\varepsilon \leq 1$ ($%
\varepsilon =\frac{E}{E_{p}}$, which $E$ and $E_{P}$ are particle and Planck
energies, respectively. See Refs. \cite%
{MagueijoII,GarattiniSar,Amelino-Camelia,Peng,Ali,HendiFEP,HendiPEM} for
more details). For $H(\varepsilon )=1$, the equation (\ref{TOVT}) yields to
the usual TOV equation \cite{TolmanI,TolmanII,Oppenheimer} (see Eq. (\ref%
{TOVEN})).

It is notable that there are known three cases for the rainbow function. The
first one is related to the hard spectra from gamma-ray bursts \cite%
{AmelinoNature}, with the following form%
\begin{equation}
F\left( \varepsilon \right) =\frac{e^{\beta \varepsilon }-1}{\beta
\varepsilon },~~~~\&~~~~H\left( \varepsilon \right) =1.
\end{equation}

The second one is motivated by studies conducted in loop quantum gravity and
non-commutative geometry \cite{Jacob,AmelinoLRR} as%
\begin{equation}
F\left( \varepsilon \right) =1,~~~~\&~~~~H\left( \varepsilon \right) =\sqrt{%
1-\eta \varepsilon ^{n}}.
\end{equation}

The third one of energy function is due to the consideration of constancy of
the velocity of light \cite{MagueijoI}%
\begin{equation}
F\left( \varepsilon \right) =H\left( \varepsilon \right) =\frac{1}{1-\lambda
\varepsilon },
\end{equation}%
where in the above models $\beta $, $\eta $\ and $\lambda $\ are constants
which can be adjusted by experiment. Here, we are going to investigate more
properties of the neutron stars by using the obtained HEE in gravity's
rainbow. As it was shown, the obtained HEE for gravity's rainbow only
depends to rainbow function of $H\left( \varepsilon \right) $. This shows
that, the first model can not be useful for investigating the structure of
neutron stars, because in this case the extracted HEE will yield Einstein's
HEE (TOV equation) and we do not see any effects of rainbow functions. We
consider the third model for three reasons; i) more effectiveness; this
model (third model) affects temporal and spatial coordinates,
simultaneously, on the contrary, the effects of the first and second models
are only on temporal coordinate and spatial coordinates, respectively. ii)
low number of free parameters; there are two free parameters for the second
model, whereas the third model has one free parameter. iii)\textbf{\ }%
phenomenologically logical; third set of rainbow functions comes from the
constancy of light velocity, which we would like to respect it. The results
are collected in table \ref{tab3}. As one can see, by increasing rainbow
function, the maximum mass and the corresponding radius of neutron star
increase. This emphasizes the contributions of the gravity's rainbow on
properties of neutron stars.

The average density of neutron star in this gravity can be written as%
\begin{equation}
\overline{\rho }=\frac{3M_{eff}}{4\pi R^{3}}.  \label{Adensity}
\end{equation}

Table \ref{tab3} shows that the average density decreases when $%
H(\varepsilon )$ increases. The Schwarzschild radius in this gravity is
given as \cite{HendiBEP}%
\begin{equation}
R_{Sch}=\frac{2GM_{eff}}{c^{2}},  \label{SchR}
\end{equation}%
the Schwarzschild radius increases when $H(\varepsilon )$ increases.

We obtain the compactness of a spherical object ($\sigma $), by using the
obtained Schwarzschild radius (Eq. (\ref{SchR})) and Eqs. (\ref{sigma}) and (%
\ref{dMass}), as%
\begin{equation}
\sigma =\frac{8\pi G}{c^{2}H^{2}(\varepsilon )R}\int_{0}^{R}r^{2}\rho \left(
r\right) dr.
\end{equation}

The gravitational redshift in gravity's rainbow can be written in the
following form \cite{HendiBEP}
\begin{equation}
z=\frac{1}{\sqrt{1-\frac{2GM_{eff}}{c^{2}R}}}-1=\frac{1}{\sqrt{1-\frac{8\pi G%
}{c^{2}H^{2}\left( \varepsilon \right) R}\int_{0}^{R}r^{2}\rho \left(
r\right) dr}}-1=\frac{1}{\sqrt{1-\sigma }}-1.
\end{equation}

The above equations show that the compactness of a spherical object and the
gravitational redshift depend on both the rainbow function ($H^{2}\left(
\varepsilon \right) $) and the radius of neutron star ($R$). On the other
hand, numerical calculations showed that increasing the rainbow function
leads to increasing the radius of neutron star, too. So, based on numerical
calculations, we have concluded that changing the rainbow function does not
have any significant role in $\sigma $\ and $z$\ up to our numeric precision
(see table. \ref{tab3}). Also, it was shown that, these stars have the
dynamical stability (see Figs. 5 and 6 of Ref. \cite{HendiBEP}, for more
details). 
\begin{table*}[tbp]
\caption{Structure properties of neutron star in gravity's rainbow with $F(%
\protect\varepsilon )=H(\protect\varepsilon )=\frac{1}{1-\protect\lambda
\protect\varepsilon }$ for $\protect\lambda =2$.}
\label{tab3}
\begin{center}
\begin{tabular}{cccccccccc}
\hline\hline
$E(E_{pl})$ & $H(\varepsilon)$ & ${M_{max}}\ (M_{\odot})$ & $R\ (km)$ & $%
R_{Sch}\ (km)$ & $\overline{\rho }$ $(10^{15}g$ $cm^{-3})$ & $\sigma(\frac{%
R_{Sch}}{R})$ & $z$ & $K(10^{-7}$ $m^{-2})$ & $\frac{4c^{2}R}{9G}\
(M_{\odot})$ \\ \hline\hline
$E \ll 1$ & $1.00$ & $1.68$ & $8.42$ & $4.95$ & $1.34$ & $0.59$ & $0.56$ & $%
0.29$ & $2.54 $ \\ \hline
$0.0833$ & $1.20$ & $2.02$ & $10.10$ & $5.95$ & $0.93$ & $0.59$ & $0.56$ & $%
0.25$ & $3.04$ \\ \hline
$0.1428$ & $1.40$ & $2.36$ & $11.79$ & $6.99$ & $0.69$ & $0.59$ & $0.56$ & $%
0.24$ & $3.55$ \\ \hline
$0.1875$ & $1.60$ & $2.69$ & $13.47$ & $7.93$ & $0.52$ & $0.59$ & $0.56$ & $%
0.24$ & $4.06$ \\ \hline
$0.2222$ & $1.80$ & $3.03$ & $15.16$ & $8.93$ & $0.41$ & $0.59$ & $0.56$ & $%
0.24$ & $4.57$ \\ \hline
$0.2500$ & $2.00$ & $3.37$ & $16.84$ & $9.93$ & $0.34$ & $0.59$ & $0.56$ & $%
0.24$ & $5.08$ \\ \hline\hline
&  &  &  &  &  &  &  &  &
\end{tabular}%
\end{center}
\end{table*}

Here we want to obtain the curvature at the surface of a neutron star in
gravity's rainbow. Considering the metric (\ref{metric}) with field equation
(\ref{fieldEq}), we can obtain the Kretschmann scalar as\textbf{\ }
\begin{equation}
K=\frac{2H(\varepsilon )}{R^{2}}\sqrt{\left[ 1-\frac{4GM\left(
c^{2}R-3GM\right) }{c^{4}R^{2}}\right] H^{4}(\varepsilon )+\frac{2\left(
2GM-c^{2}R\right) H^{2}(\varepsilon )}{c^{2}R}+1}.  \label{Krain}
\end{equation}

It is notable that when $H(\varepsilon )=1$, the Kretschmann scalar of
Einstein gravity (Eq. \ref{K}) will be recovered. The results show that, by
increasing rainbow function which leads to increasing of radius and mass of
neutron stars, but the strength of gravity decreases.

Another important quantity which is interesting for evaluation is related to
the upper mass limit of a static spherical neutron star with uniform density
in gravity's rainbow. For this purpose, we have to obtain Buchdahl theorem
in an energy dependent spacetime. Solving the integral given in Eq. (\ref%
{dMass}), we obtain\textbf{\ }%
\begin{equation}
M_{eff}=\frac{4\pi \rho }{3H^{2}(\varepsilon )}r^{3}.  \label{Mass}
\end{equation}

Inserting Eq. (\ref{Mass}) into the obtained HEE in gravity's rainbow (\ref%
{TOVT}), we have\textbf{\ }%
\begin{equation}
\frac{dP}{dr}=-\frac{4\pi Gr}{3c^{4}H^{2}(\varepsilon )}\frac{\left(
c^{2}\rho +3P\right) \left( c^{2}\rho +P\right) }{\left( 1-\frac{8\pi G}{%
3c^{2}H^{2}(\varepsilon )}\rho r^{2}\right) }.
\end{equation}

Integrating from a central pressure $p_{c}=p(r=0)$, we have\textbf{\ }%
\begin{equation}
\int_{p_{c}}^{p}\frac{dP^{\prime }}{\left( c^{2}\rho +3P^{\prime }\right)
\left( c^{2}\rho +P^{\prime }\right) }=-\frac{4\pi G}{3c^{4}H^{2}(%
\varepsilon )}\int_{0}^{r}\frac{r^{\prime }dr^{\prime }}{\left( 1-\frac{8\pi
G}{3c^{2}H^{2}(\varepsilon )}\rho r^{\prime 2}\right) },  \label{int2}
\end{equation}%
\textbf{\ }in which the integral on the right-hand side of the above
equation is%
\begin{equation*}
-\frac{4\pi G}{3c^{4}H^{2}(\varepsilon )}\int_{0}^{r}\frac{r^{\prime
}dr^{\prime }}{\left( 1-\frac{8\pi G}{3c^{2}H^{2}(\varepsilon )}\rho
r^{\prime 2}\right) }=\frac{1}{4c^{2}\rho }\int_{1}^{1-\frac{8\pi G}{%
3c^{2}H^{2}(\varepsilon )}\rho r^{2}}\frac{dy}{y}=\frac{1}{4c^{2}\rho }\ln
y\left\vert _{1}^{1-\frac{8\pi G}{3c^{2}H^{2}(\varepsilon )}\rho
r^{2}}\right.
\end{equation*}%
\textbf{\ \ }%
\begin{equation}
=\frac{1}{4c^{2}\rho }\ln \left( 1-\frac{8\pi G}{3c^{2}H^{2}(\varepsilon )}%
\rho r^{2}\right) ,  \label{left}
\end{equation}%
which we have used of $y=1-\frac{8\pi G}{3H^{2}(\varepsilon )}r^{\prime
2}\rho $\ in the above equation. On the other hand, the left-hand side of
Eq. (\ref{int2}) is%
\begin{equation*}
\int_{p_{c}}^{p}\frac{dP^{\prime }}{\left( c^{2}\rho +3P^{\prime }\right)
\left( c^{2}\rho +P^{\prime }\right) }=\frac{1}{2c^{2}\rho }\left. \ln
\left( \frac{c^{2}\rho +3P^{\prime }}{c^{2}\rho +P^{\prime }}\right)
\right\vert _{p_{c}}^{p}
\end{equation*}%
\textbf{\ \ }%
\begin{equation}
=\frac{1}{2c^{2}\rho }\left[ \ln \left( \frac{c^{2}\rho +3P}{c^{2}\rho +P}%
\right) -\ln \left( \frac{c^{2}\rho +3P_{c}}{c^{2}\rho +P_{c}}\right) \right]
.  \label{right}
\end{equation}%
\textbf{\ }

Combining Eqs. (\ref{left}) and (\ref{right}), we can write\textbf{\ }%
\begin{equation}
\frac{1}{2c^{2}\rho }\left[ \ln \left( \frac{c^{2}\rho +3P}{c^{2}\rho +P}%
\right) -\ln \left( \frac{c^{2}\rho +3P_{c}}{c^{2}\rho +P_{c}}\right) \right]
=\frac{1}{4c^{2}\rho }\ln \left( 1-\frac{8\pi G}{3c^{2}H^{2}(\varepsilon )}%
\rho r^{2}\right) ,
\end{equation}%
so we have%
\begin{equation}
\frac{c^{2}\rho +3P}{c^{2}\rho +P}=\frac{c^{2}\rho +3P_{c}}{c^{2}\rho +P_{c}}%
\sqrt{1-\frac{8\pi G}{3c^{2}H^{2}(\varepsilon )}\rho r^{2}},
\end{equation}%
in which by using Eq. (\ref{Mass}), we can rewrite it as\textbf{\ }%
\begin{equation}
\frac{c^{2}\rho +3P}{c^{2}\rho +P}=\frac{c^{2}\rho +3P_{c}}{c^{2}\rho +P_{c}}%
\sqrt{1-\frac{2GM_{eff}\left( r,\varepsilon \right) }{c^{2}r}}.
\end{equation}

In order to obtain an expression for finding how the radius of star depends
on its mass and density, we use the fact that the pressure of a star on its
surface vanishes, i.e., $p(r)=0$ for $r=R$ ($p(R)=0$). So we can obtain%
\textbf{\ }%
\begin{equation}
\frac{2GM_{eff}}{c^{2}R}=1-\left( \frac{c^{2}\rho +P_{c}}{c^{2}\rho +3P_{c}}%
\right) ^{2}.  \label{ew}
\end{equation}

Introducing a new dimensionless variable as $x=\frac{c^{2}\rho }{P_{c}}$\ ,
we can rewrite Eq. (\ref{ew}) with the following form\textbf{\ }%
\begin{equation}
\frac{2GM_{eff}}{c^{2}R}=1-\left( \frac{x+1}{x+3}\right) ^{2}=1-\left[ S(x)%
\right] ^{2},  \label{ew2}
\end{equation}%
where $S(x)=\frac{x+1}{x+3}$. According to this fact the variable $x$\ is
functions of $P_{c}$\ and $\rho $\ in which both them can never be negative,
hence $x$\ can not be negative and also due to the negative sign of the $%
S(x) $, Eq. (\ref{ew2}) reaches its maximum value when $S(x)$\ reaches its
minimum \ value. In order to find the minimum of the $S(x)$, we find the
derivative as $\frac{dS(x)}{dx}=\frac{2}{\left( x+3\right) ^{2}}$. This is
always positive. Thus $S(x)$\ increases with increasing $x$, and the minimum
of $S(x)$\ is located at $x=0$. Inserting $x=0$\ in Eq. (\ref{ew2}), we have%
\textbf{\ \ }%
\begin{equation}
\frac{GM_{eff\max }}{c^{2}R}=\frac{4}{9},~~~~or~~~~~M_{eff\max }=\frac{%
4c^{2}R}{9G}.  \label{max}
\end{equation}

The equation (\ref{max}) gives the value for the maximum mass of a star of
given radius in gravity's rainbow. As a results of our calculation for
finding the upper mass limit for a static spherical neutron star with
uniform density in gravity's rainbow, the obtained mass of neutron stars we
must be in the range\textbf{\ \ }%
\begin{equation}
M_{eff}<\frac{4c^{2}R}{9G}.  \label{rang}
\end{equation}

It is notable that, for $H(\varepsilon )=1$, this limit reduces to Einstein
case ($M<\frac{4c^{2}R}{9G}$). Our results of mass of magnetic neutron stars
in gravity's rainbow show that these masses are in the obtained range of Eq.
(\ref{rang}).

The maximum mass and radius of neutron star are increasing functions of the
rainbow function while the average density and the Kretschmann scalar were a
decreasing function of it. These properties indicate that increasing the
rainbow function results into expansion of the neuron star.

\section{Structure properties of magnetized neutron stars in gravity's
rainbow}

\label{STRUCTURE}

\begin{figure*}[tbp]
$%
\begin{array}{cc}
\epsfxsize=8cm \epsffile{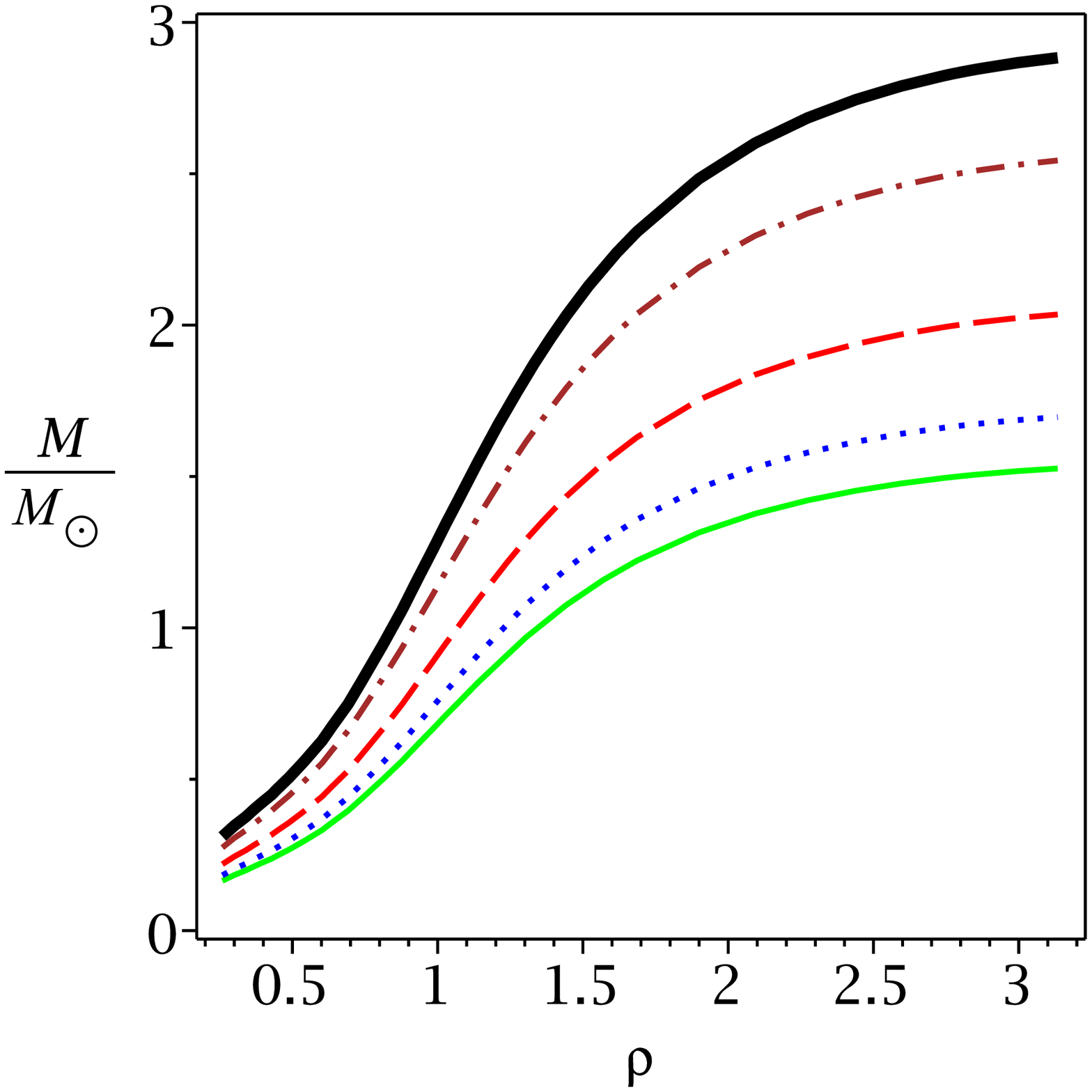} & \epsfxsize=8cm %
\epsffile{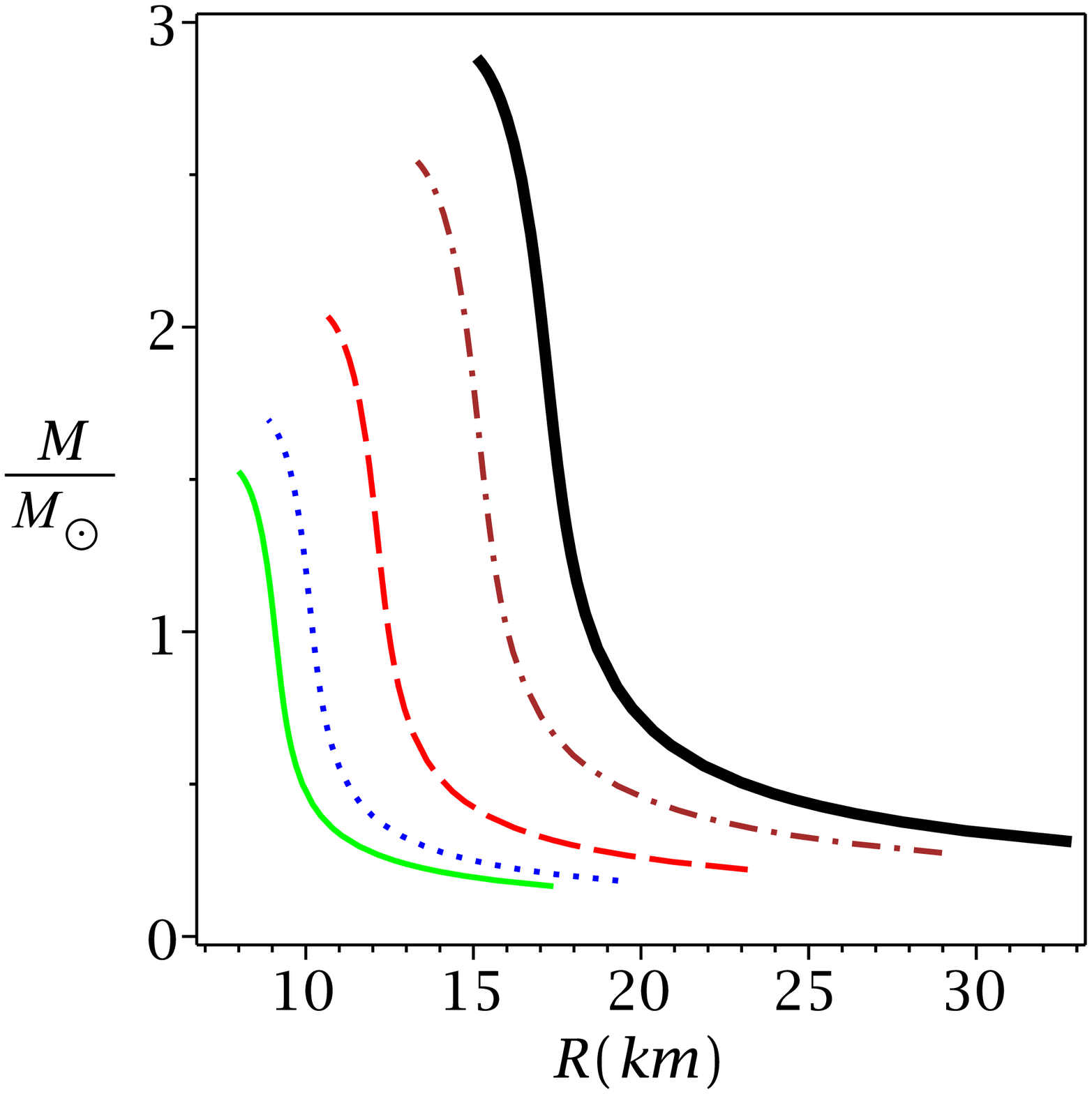}%
\end{array}
$%
\caption{Gravitational mass versus the density $(10^{15}g$ $cm^{-3})$ (right
diagram) and radius (left diagram), for $B=5\times 10^{18}G$, $H(\protect%
\varepsilon )=0.9$ (continuous line), $H(\protect\varepsilon )=1$ (doted
line), $H(\protect\varepsilon )=1.2$ (dashed line), $H(\protect\varepsilon %
)=1.5$ (dashed-dotted line), and $H(\protect\varepsilon )=1.7$ (bold line).}
\label{Fig4}
\end{figure*}

Considering the EoS introduced in Appendix A and B, we want to calculate the
properties of magnetic neutron star in gravity's rainbow in this section.
The effects of rainbow function and magnetic field on the diagrams related
to $M-\rho $ and $M-R$ relations are presented in Figs. \ref{Fig4} and \ref%
{Fig5}. As one can see, by increasing rainbow function and magnetic field,
the maximum mass and radius of magnetized neutron stars increase. It is
notable that for $H(\varepsilon )=1$, the maximum mass of these stars
reduces to the results obtained in the Einstein gravity (see table \ref{tab2}%
), as expected. By increasing rainbow function ($H(\varepsilon )>1$) and
magnetic field, the maximum mass of these stars will be larger than the case
of the Einstein gravity (see table \ref{tab4} and Figs. \ref{Fig4} and \ref%
{Fig5}). This emphasizes the contributions of the gravity's rainbow and
magnetic field on properties of neutron stars. The effects of magnetic
fields on the gravitational masses of neutron stars and M-R relation for a
fixed amount of the rainbow function are presented in Fig. \ref{Fig6}. The
maximum mass and radius of magnetized neutron star for each value of the
rainbow function increase when the magnetic field grows (see table \ref{tab4}%
). Our results cover the mass measurement of massive magnetized neutron
stars. For example; about $1.8M_{\odot }$ for Vela X-1 \cite{Rawlset}, PSR
J1614-2230 \cite{Demorest} about $1.97M_{\odot }$, PSR J0348+0432 \cite%
{Antoniadis} about $2.01M_{\odot }$, 4U 1700-377 \cite{Clark et al} about $%
2.4M_{\odot }$, and J1748-2021B \cite{Freire} about $2.7M_{\odot }$.

Previously, Rhoades and Ruffini have shown that by employing the principles
of causality and Le Chantelier in Einstein gravity, the mass of neutron star
can not be larger than $3.2M_{\odot }$ \cite{Ruffini}. Here, it was shown
that the principles of causality and Le Chantelier are valid for the EoS in
this paper (see Appendix B). In addition, we find that depending on choices
of rainbow function, the mass of the neutron star could be larger than $%
3.2M_{\odot }$.

The results for the Schwarzschild radius show that for each value of the
magnetic field with stronger gravity's rainbow, the Schwarzschild radius has
higher value. In addition, with a fixed amount of rainbow function, the
Schwarzschild radius grows when the magnetic field increases. Moreover, the
average density of the magnetized neutron star shows that the average
density decreases when both $H(\varepsilon )$ and $B$ increase.

The compactness of this star confirms when $H(\varepsilon )$ increases, the
strength of gravity is the same, but by increasing the magnetic field, the
strength of gravity decreases and this may increase the radius of this star.

Using Eq. (\ref{Krain}), we find that for a fixed value of the magnetic
field by increasing rainbow function, the strength of gravity of a neutron
star has smaller values. In addition, with a fixed amount of rainbow
function, the Kretschmann scalar decreases when the magnetic field increases.

It is notable that, by increasing $B$ leads to decreasing the gravitational
redshift. On the other hand, considering a fixed value of magnetic field,
the rainbow function does not affect the gravitational redshift. Our results
show that, the maximum redshift for this star is about $0.51$, which is
lower than the upper bound on the surface redshift for a subluminal equation
of state, i.e $z=0.85$ \cite{HaeselII}.

In order to investigate the dynamical stability of magnetized neutron star,
we plot the adiabatic index versus the radius in Fig. \ref{Fig7}. As one can
see, these stars have the dynamical stability inside.

\begin{table*}[tbp]
\caption{Properties of magnetic neutron star in gravity's rainbow with $F(
\protect\varepsilon) =H( \protect\varepsilon) =\frac{1}{1-\protect\lambda
\protect\varepsilon }$ for $\protect\lambda =2$.}
\label{tab4}
\begin{center}
\begin{tabular}{ccccccccccc}
\hline\hline
$E(E_{pl})$ & $H(\varepsilon)$ & $B(G)$ & ${M_{max}}\ (M_{\odot})$ & $R\
(km) $ & $R_{Sch}\ (km)$ & $\overline{\rho }$ $(10^{15}g$ $cm^{-3})$ & $%
\sigma (\frac{R_{Sch}}{R})$ & $z$ & $K(10^{-7}$ $m^{-2})$ & $\frac{4c^{2}R}{%
9G}\ (M_{\odot})$ \\ \hline\hline
$E \ll 1$ & $1.00$ & $%
\begin{array}{c}
5\times 10^{18} \\
1\times 10^{19}%
\end{array}
$ & $%
\begin{array}{c}
1.70 \\
1.71%
\end{array}
$ & $%
\begin{array}{c}
8.88 \\
9.36%
\end{array}
$ & $%
\begin{array}{c}
5.01 \\
5.04%
\end{array}
$ & $%
\begin{array}{c}
1.15 \\
0.99%
\end{array}
$ & $%
\begin{array}{c}
0.56 \\
0.54%
\end{array}
$ & $%
\begin{array}{c}
0.51 \\
0.47%
\end{array}
$ & $%
\begin{array}{c}
0.25 \\
0.21%
\end{array}
$ & $%
\begin{array}{c}
2.68 \\
2.82%
\end{array}
$ \\ \hline
$0.0833$ & $1.20$ & $%
\begin{array}{c}
5\times 10^{18} \\
1\times 10^{19}%
\end{array}
$ & $%
\begin{array}{c}
2.03 \\
2.05%
\end{array}
$ & $%
\begin{array}{c}
10.66 \\
11.23%
\end{array}
$ & $%
\begin{array}{c}
5.98 \\
6.04%
\end{array}
$ & $%
\begin{array}{c}
0.79 \\
0.69%
\end{array}
$ & $%
\begin{array}{c}
0.56 \\
0.54%
\end{array}
$ & $%
\begin{array}{c}
0.51 \\
0.47%
\end{array}
$ & $%
\begin{array}{c}
0.21 \\
0.18%
\end{array}
$ & $%
\begin{array}{c}
3.21 \\
3.38%
\end{array}
$ \\ \hline
$0.1428$ & $1.40$ & $%
\begin{array}{c}
5\times 10^{18} \\
1\times 10^{19}%
\end{array}
$ & $%
\begin{array}{c}
2.37 \\
2.40%
\end{array}
$ & $%
\begin{array}{c}
12.43 \\
13.10%
\end{array}
$ & $%
\begin{array}{c}
6.99 \\
7.07%
\end{array}
$ & $%
\begin{array}{c}
0.58 \\
0.51%
\end{array}
$ & $%
\begin{array}{c}
0.56 \\
0.54%
\end{array}
$ & $%
\begin{array}{c}
0.51 \\
0.47%
\end{array}
$ & $%
\begin{array}{c}
0.20 \\
0.18%
\end{array}
$ & $%
\begin{array}{c}
3.75 \\
3.95%
\end{array}
$ \\ \hline
$0.1875$ & $1.60$ & $%
\begin{array}{c}
5\times 10^{18} \\
1\times 10^{19}%
\end{array}
$ & $%
\begin{array}{c}
2.71 \\
2.74%
\end{array}
$ & $%
\begin{array}{c}
14.21 \\
14.98%
\end{array}
$ & $%
\begin{array}{c}
7.99 \\
8.08%
\end{array}
$ & $%
\begin{array}{c}
0.45 \\
0.39%
\end{array}
$ & $%
\begin{array}{c}
0.56 \\
0.54%
\end{array}
$ & $%
\begin{array}{c}
0.51 \\
0.47%
\end{array}
$ & $%
\begin{array}{c}
0.20 \\
0.18%
\end{array}
$ & $%
\begin{array}{c}
4.28 \\
4.52%
\end{array}
$ \\ \hline
$0.2222$ & $1.80$ & $%
\begin{array}{c}
5\times 10^{18} \\
1\times 10^{19}%
\end{array}
$ & $%
\begin{array}{c}
3.05 \\
3.08%
\end{array}
$ & $%
\begin{array}{c}
16.10 \\
16.84%
\end{array}
$ & $%
\begin{array}{c}
8.99 \\
9.08%
\end{array}
$ & $%
\begin{array}{c}
0.35 \\
0.31%
\end{array}
$ & $%
\begin{array}{c}
0.56 \\
0.54%
\end{array}
$ & $%
\begin{array}{c}
0.51 \\
0.47%
\end{array}%
$ & $%
\begin{array}{c}
0.20 \\
0.18%
\end{array}
$ & $%
\begin{array}{c}
4.85 \\
5.08%
\end{array}
$ \\ \hline
$0.2500$ & $2.00$ & $%
\begin{array}{c}
5\times 10^{18} \\
1\times 10^{19}%
\end{array}
$ & $%
\begin{array}{c}
3.39 \\
3.42%
\end{array}
$ & $%
\begin{array}{c}
17.76 \\
18.72%
\end{array}
$ & $%
\begin{array}{c}
9.99 \\
10.08%
\end{array}
$ & $%
\begin{array}{c}
0.29 \\
0.25%
\end{array}
$ & $%
\begin{array}{c}
0.56 \\
0.54%
\end{array}
$ & $%
\begin{array}{c}
0.51 \\
0.47%
\end{array}
$ & $%
\begin{array}{c}
0.20 \\
0.18%
\end{array}
$ & $%
\begin{array}{c}
5.35 \\
5.64%
\end{array}
$ \\ \hline
&  &  &  &  &  &  &  &  &  &
\end{tabular}%
\end{center}
\end{table*}
\begin{figure*}[tbp]
$%
\begin{array}{cc}
\epsfxsize=8cm \epsffile{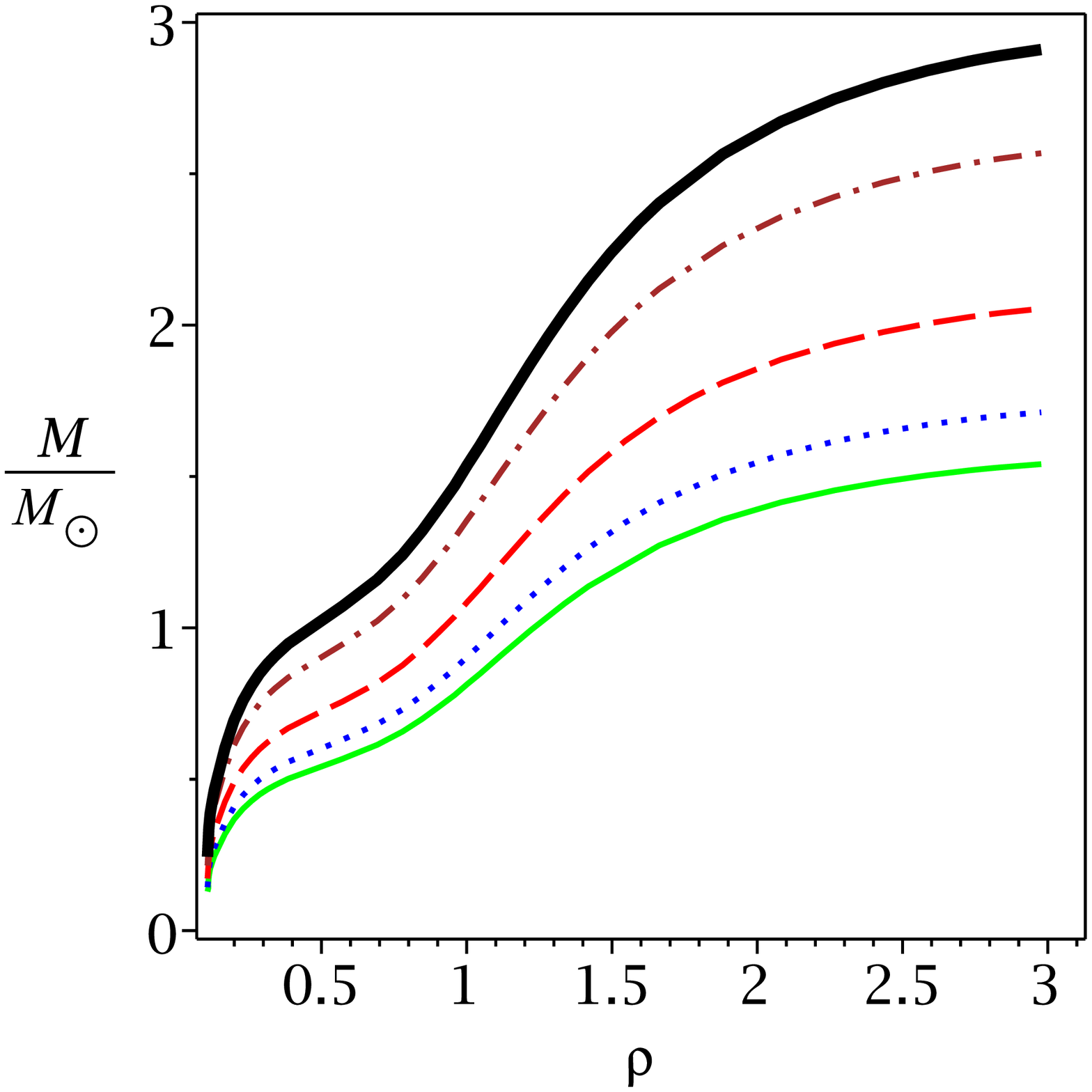} & \epsfxsize=8cm %
\epsffile{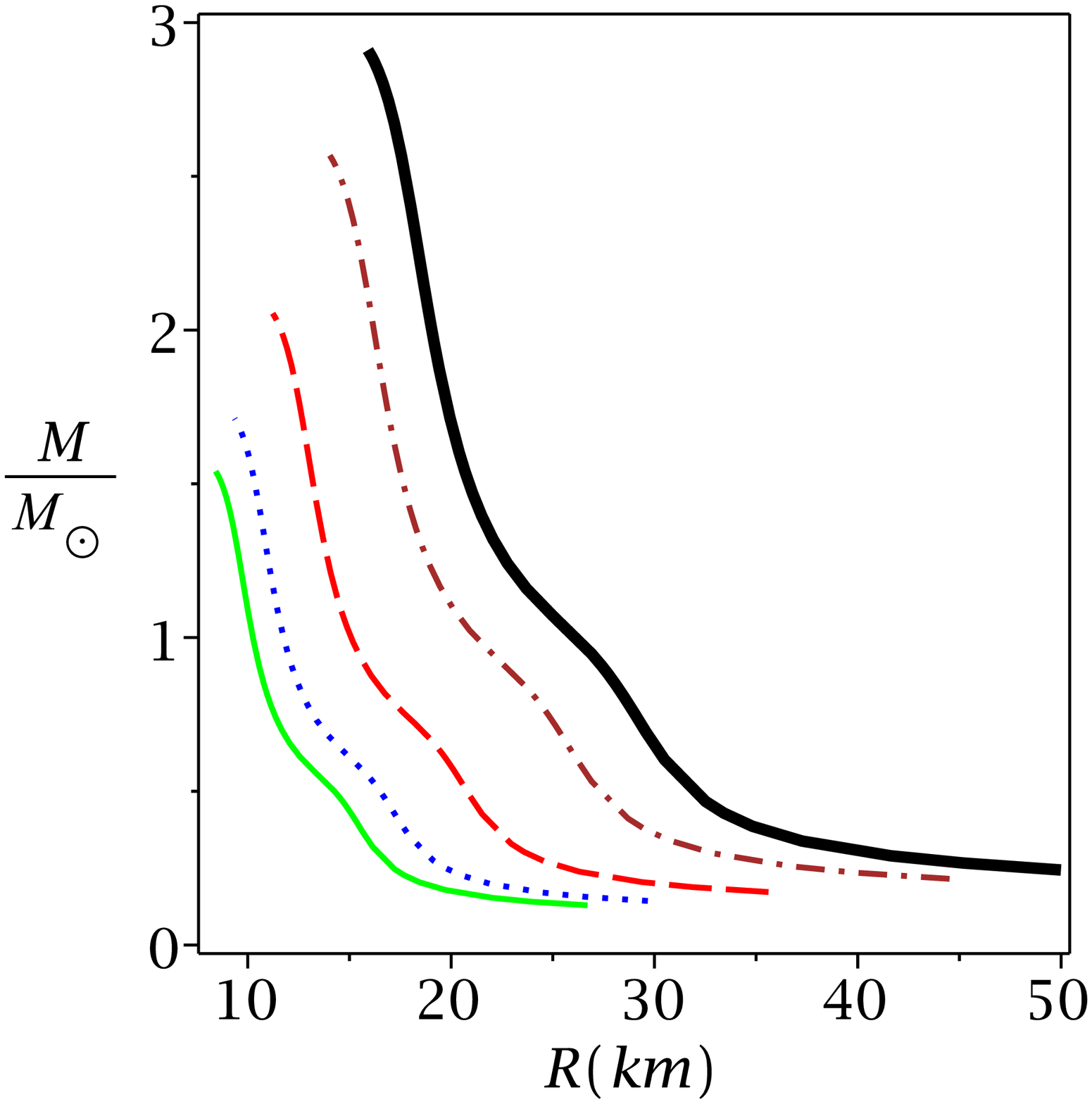}%
\end{array}
$%
\caption{Same as Figs. \protect\ref{Fig4} but for $B=1\times 10^{19}G$.}
\label{Fig5}
\end{figure*}

\begin{figure*}[tbp]
$%
\begin{array}{cc}
\epsfxsize=8cm \epsffile{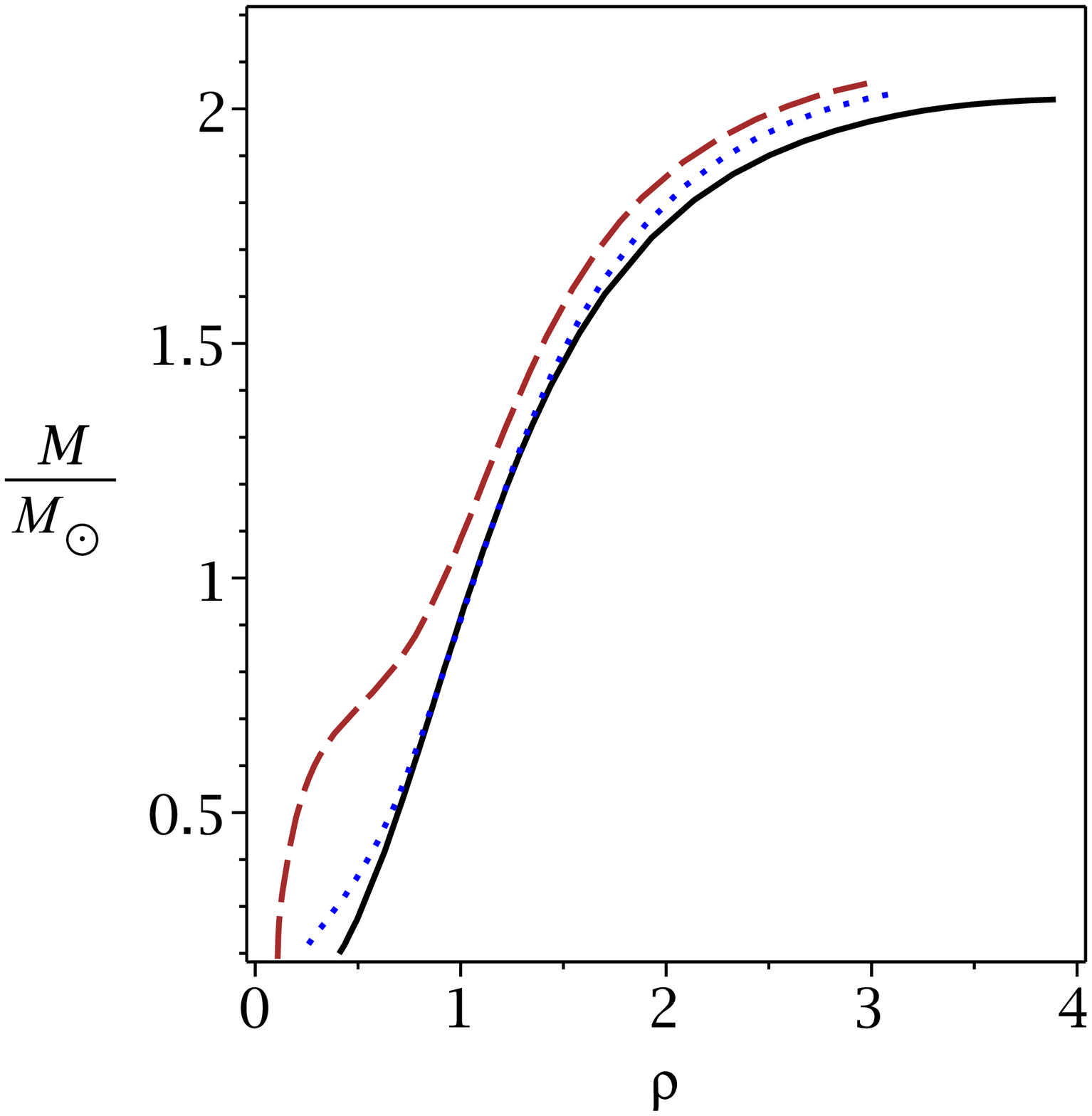} & \epsfxsize=8cm \epsffile{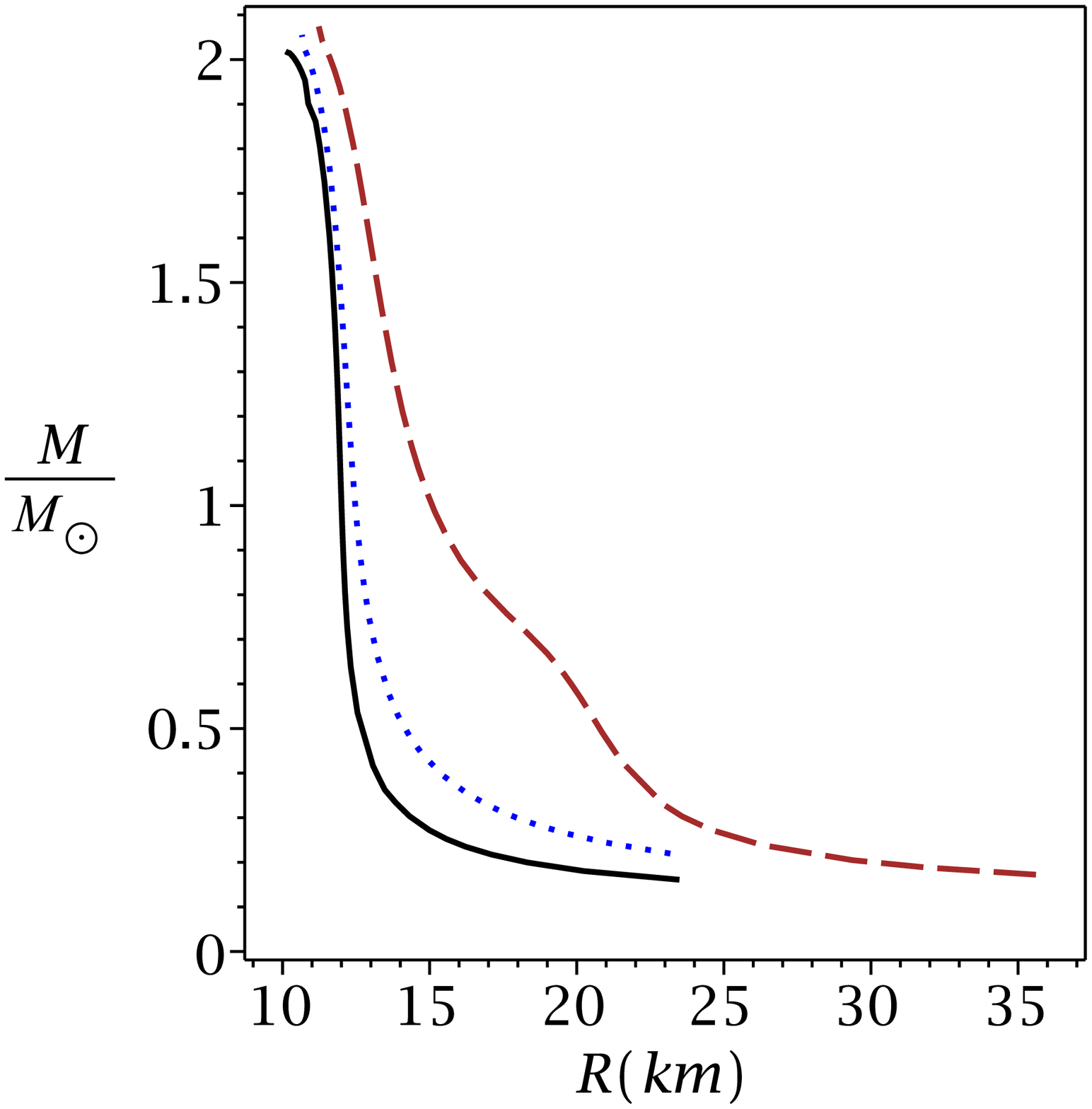}%
\end{array}
$%
\caption{Gravitational mass versus the density $(10^{15}g$ $cm^{-3})$ (right
diagram) and radius (left diagram), for $H(\protect\varepsilon )=1.2$, $B=0$
(continuous line), $B=5\times 10^{18}G$ (doted line), and $B=1\times
10^{19}G $ (dashed line).}
\label{Fig6}
\end{figure*}

\begin{figure}[tbp]
\epsfxsize=12cm \centerline{\epsffile{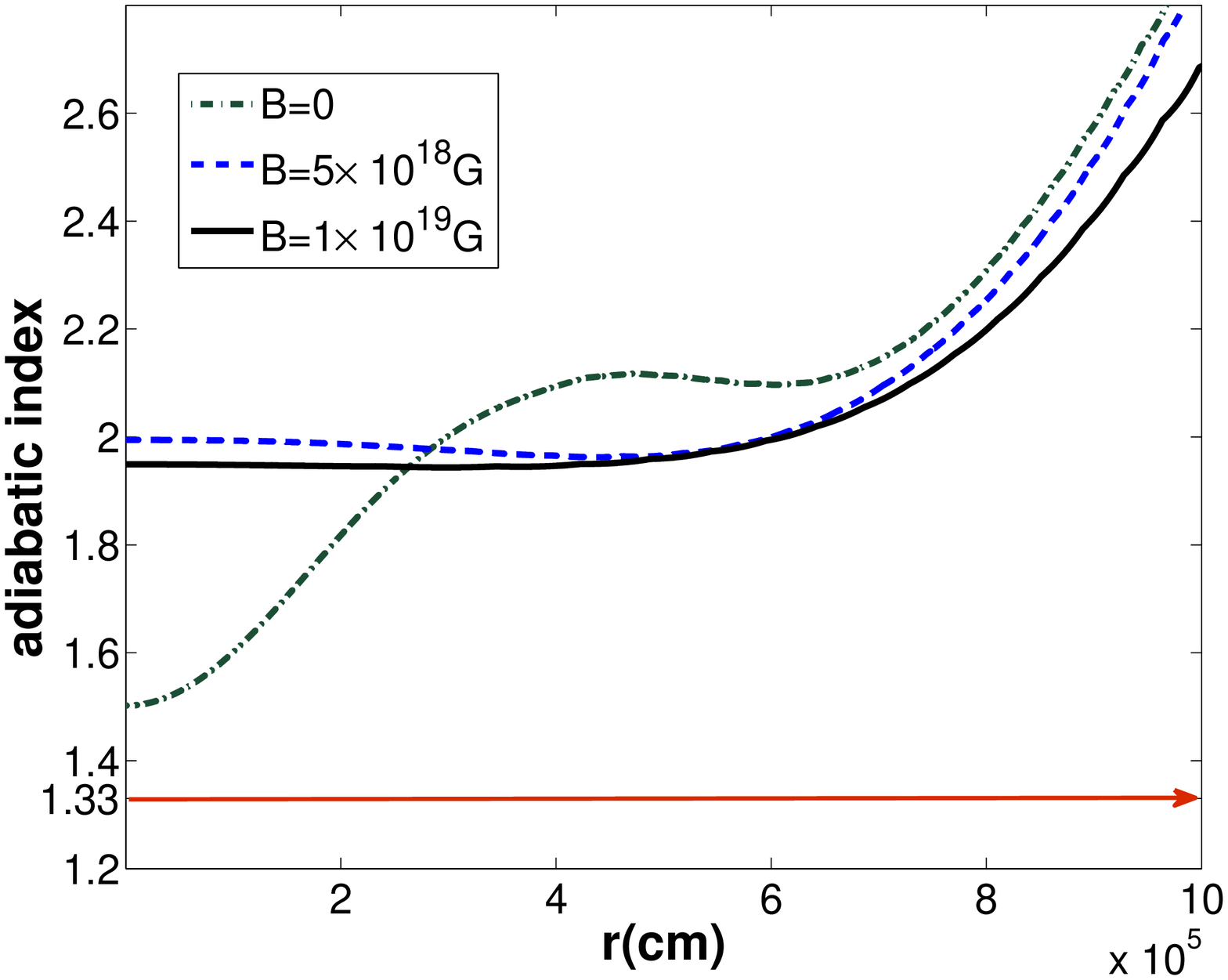}}
\caption{The adiabatic index versus the radius for $H(\protect\varepsilon %
)=1.4$.}
\label{Fig7}
\end{figure}


As previously mentioned, the pressure can be written as $P_{B}\propto B^{2}$%
. In other words, by adding the magnetic pressure ($P_{B}$)\ to the pressure
of the fluid ($P$), the pressure power increases. On the other hand, in
order to have a stable magnetic neutron star, the gravitational force has to
increase and this leads to increasing of mass of magnetic neutron star.

\section{Comparison between theory and observations}

\label{Comparison}

One of the interesting features of our calculations is related to comparison
of the theory and its predictions with the empirical evidences. For this
purpose, we compare our results with the observational data. We have
presented this comparison in table \ref{tab5}. It is clear that the results
derived in this theory are in agreement with the results obtained through
observations. It is notable that for non-magnetized neutron stars, the
radius obtained through gravity's rainbow are smaller than the observed
values. However, the radius of the magnetized neutron stars are in the range
of the observed values. This is due to the fact that we have considered the
magnetic field of neutron stars in our calculations (see table \ref{tab5}
for more details).

\begin{table*}[tbp]
\caption{Mass and radius of neutron stars through observations and theory by
considering $F( \protect\varepsilon) =H( \protect\varepsilon) =\frac{1}{1-%
\protect\lambda \protect\varepsilon }$, and $\protect\lambda =2$.}
\label{tab5}
\begin{center}
\begin{tabular}{cc}
\begin{tabular}{ccccccccccc}
\hline\hline
& $Observations$ &  &  &  &  &  & $Theory$ &  &  &  \\
$name$ & ${M}\ (M_{\odot})$ & $R\ (km)$ &  &  & $B(G)$ & $E(E_{pl})$ & $%
H(\varepsilon)$ & ${M}\ (M_{\odot})$ & $R\ (km)$ &  \\ \hline\hline
$J1748-2021B$ & $2.70$ & $unknown$ &  &  & $%
\begin{array}{c}
0.0 \\
5\times 10^{18} \\
1\times 10^{19}%
\end{array}
$ & $%
\begin{array}{c}
0.1875 \\
0.1855 \\
0.1835%
\end{array}
$ & $%
\begin{array}{c}
1.60 \\
1.59 \\
1.58%
\end{array}
$ & $2.70$ & $%
\begin{array}{c}
13.52 \\
14.12 \\
14.78%
\end{array}
$ &  \\ \hline
$4U\ 1700-377$ & $2.40$ & $unknown$ &  &  & $%
\begin{array}{c}
0.0 \\
5\times 10^{18} \\
1\times 10^{19}%
\end{array}
$ & $%
\begin{array}{c}
0.1478 \\
0.1453 \\
0.1428%
\end{array}
$ & $%
\begin{array}{c}
1.42 \\
1.41 \\
1.40%
\end{array}
$ & $2.40$ & $%
\begin{array}{c}
12.00 \\
12.57 \\
13.15%
\end{array}%
$ &  \\ \hline
$PSR\ J0348+0432$ & $2.01$ & $13(\pm 2)$ &  &  & $%
\begin{array}{c}
0.0 \\
5\times 10^{18} \\
1\times 10^{19}%
\end{array}
$ & $%
\begin{array}{c}
0.0798 \\
0.0762 \\
0.0726%
\end{array}
$ & $%
\begin{array}{c}
1.19 \\
1.18 \\
1.17%
\end{array}
$ & $2.01$ & $%
\begin{array}{c}
10.01 \\
10.54 \\
11.00%
\end{array}
$ &  \\ \hline
$PSR\ J1614-2230$ & $1.97$ & $13(\pm 2)$ &  &  & $%
\begin{array}{c}
0.0 \\
5\times 10^{18} \\
1\times 10^{19}%
\end{array}
$ & $%
\begin{array}{c}
0.0726 \\
0.0689 \\
0.0652%
\end{array}
$ & $%
\begin{array}{c}
1.17 \\
1.16 \\
1.15%
\end{array}
$ & $1.97$ & $%
\begin{array}{c}
9.88 \\
10.35 \\
10.80%
\end{array}%
$ &  \\ \hline
$4U\ 1608-52$ & $1.74$ & $9.3(\pm 1)$ &  &  & $%
\begin{array}{c}
0.0 \\
5\times 10^{18} \\
1\times 10^{19}%
\end{array}
$ & $%
\begin{array}{c}
0.0145 \\
0.0098 \\
0.0049%
\end{array}
$ & $%
\begin{array}{c}
1.03 \\
1.02 \\
1.01%
\end{array}%
$ & $1.74$ & $%
\begin{array}{c}
8.71 \\
9.13 \\
9.52%
\end{array}%
$ &  \\ \hline
&  &  &  &  &  &  &  &  &  &
\end{tabular}
&
\end{tabular}%
\end{center}
\end{table*}
It is worthwhile to mention that obtaining radius with the observational
measurement has specific difficulties. However, with using this theory, we
can obtain the radius of neutron stars. The results of the theoretical
calculations are presented in table \ref{tab5}. For observational cases, the
theory under consideration with strong magnetic field predicts a set of
radii for some cases such as J1748-2021B and 4U 1700-377, which are about $%
14.78$ (km) and $13.15$ (km), respectively.

\section{Conclusions}

\label{Conclusions}

The paper at hand studied the structure of magnetized neutron stars through
the use of LOCV method and $AV_{18}$ potential employing the hydrostatic
equilibrium equation in Einstein gravity. It was shown that while the
maximum mass and radius of neutron star are increasing functions of the
magnetic field, the average density, compactness, redshift and Kretschmann
scalar are decreasing functions of it. This results into expansion of the
neutron star in case of increasing power of the magnetic field. In some
senses, the magnetic field may have repulsive force property.

Next, the effects of the gravity's rainbow were explored. It was shown that
the compactness and the gravitational redshift are not sensitive to
variation of the rainbow function, the maximum mass and radius of neutron
star were increasing functions of it and the average density and Kretschmann
scalar were decreasing functions of it. Therefore, increasing the rainbow
function leads to expansion of the neutron stars.\ Furthermore, it was also
pointed out that it is possible to have magnetized neutron stars with mass
larger than $3.2M_{\odot }$ in gravity's rainbow.

Finally, a comparison with the observational data was done and it was shown
that our theoretical results are in agreement with the empirical evidences
of neutron stars.

Briefly, we obtained the quite interesting results for magnetic neutron star
in an energy dependent spacetime such as: I)\ EoS derived from microscopic
calculations satisfied Le Chatelier's principle and also both energy and
stability conditions, simultaneously (see Appendix B). II) The maximum mass
and radius of neutron star were increasing functions of magnetic field and
rainbow function. III) The obtained maximum mass of magnetic neutron star
can be more than $3.2M_{\odot }$, due to considering an energy dependent
spacetime. IV) Magnetic neutron star in gravity's rainbow was dynamically
stable. V) Obtained results of magnetic neutron stars in gravity's rainbow
are consistent with all the measured masses of pulsars and neutron stars
such as; Vela X-1 (about $1.8M_{\odot }$\ ) \cite{Rawlset}, PSR J1614-2230
(about $1.97M_{\odot }$\ ) \cite{Demorest}, PSR J0348+0432 (about $%
2.01M_{\odot }$) \cite{Antoniadis}, 4U 1700-377 (about $2.4M_{\odot }$\ )
\cite{Clark et al} and J1748-2021B (about $2.7M_{\odot }$\ ) \cite{Freire}.
VI) We have shown the magnetic field and rainbow function may have repulsive
force property. VII) We obtained the Kretschmann scalar in gravity's rainbow
as a characteristic of gravity strength and evaluated the its behavior.
VIII) We have extracted a new relation for the maximum mass of a star of
given radius in gravity's rainbow as $M_{eff}<\frac{4c^{2}R}{9G}$.

It will be worthwhile to study the effects of magnetic field on properties
of other compact objects such as quark star and white dwarf in the context
of gravity's rainbow. In this paper we considered a neutron star in static
spherical symmetric spacetime and studied simultaneously the effects of
magnetic field and rainbow function on its properties. Generalization of
static compact objects to anisotropic \cite%
{anisoI,anisoIII,anisoIV,anisoV,anisoVI}, rotating \cite%
{rotI,rotIII,rotVI,rotVII,rotVIII,rotIX,rotXI}, rapidly rotating \cite%
{rapidI,rapidII,rapidIII,rapidV,rapidVII} of compact objects and
investigating the effect of magnetic field in gravity's rainbow can be
interesting topics. We leave these issues for the future works.

\acknowledgments

We wish to thank Shiraz University Research Council. This work has been
supported financially by the Research Institute for Astronomy and
Astrophysics of Maragha, Iran.

\begin{center}
\textbf{Appendix: A LOCV formalism for magnetized neutron matter }
\end{center}

We suppose a pure homogeneous system composed of spin-up $(+)$ and spin-down
$(-)$ neutrons with the number densities of $\rho ^{(+)}$ and $\rho ^{(-)}$,
respectively. We denote the spin polarization parameter by $\delta =\frac{%
\rho ^{(+)}-\rho ^{(-)}}{\rho }$ and the total density of system by $\rho
=\rho ^{(+)}+\rho ^{(-)}$. We use LOCV method to calculate the energy of our
system \cite{BordbarRM,Rezaei}. Although in order to obtain accurate results
we should regard temperature of our system, here we consider zero
temperature for simplicity. A trial many-body wave function of the form $%
\psi =F\phi $ is considered where $\phi $ is the uncorrelated ground-state
wave function of $N$ independent neutrons, and $F$ is a proper $N$-body
correlation function. Applying Jastrow approximation \cite{Clark}, $F$ is
replaced by $F=S\prod_{i>j}f(ij)$, where $S$ is a symmetrizing operator. In
addition, a cluster expansion of the energy functional up to the two-body
term, $E([f])=\frac{1}{N}\frac{\langle \psi |H|\psi \rangle }{\langle \psi
|\psi \rangle }=E_{1}+E_{2}$, is considered. Taking the uniform magnetic
field along the $z$ direction, $B=B\widehat{k}$, the spin up and down
particles correspond to parallel and antiparallel spins with respect to the
magnetic field. The energy per particle up to the two-body term is
\begin{equation*}
E([f])=E_{1}+E_{2}-\mu _{n}B\delta ,
\end{equation*}%
where $E_{1}=\sum_{i=+,-}\frac{3}{5}\frac{\hbar ^{2}k_{F}^{(i)^{2}}}{2m}%
\frac{\rho ^{(i)}}{\rho }$ and $E_{2}=\frac{1}{2N}\sum_{ij}\langle ij|\nu
(12)|ij-ji\rangle $ are one-body and two-body energy terms, respectively.
The operator $\nu (12)$ is nuclear potential which has been given in Ref.
\cite{BordbarRM}, and $\mu _{n}=-1.9130427(5)$ is the neutron magnetic
moment (in units of the nuclear magneton). In the next step, we minimize the
two-body energy with respect to the variations in the trial many-body wave
function subject to a normalization constraint presented in Ref. \cite%
{BordbarRM}. The result of minimization of the two-body cluster energy is a
set of differential equations \cite{BordbarRM}. Solving the differential
equations leads to the two-body energy of the system.

\begin{center}
\textbf{Appendix: B EoS of magnetized neutron star matter}
\end{center}

\begin{figure*}[tbp]
$%
\begin{array}{cc}
\epsfxsize=8cm \epsffile{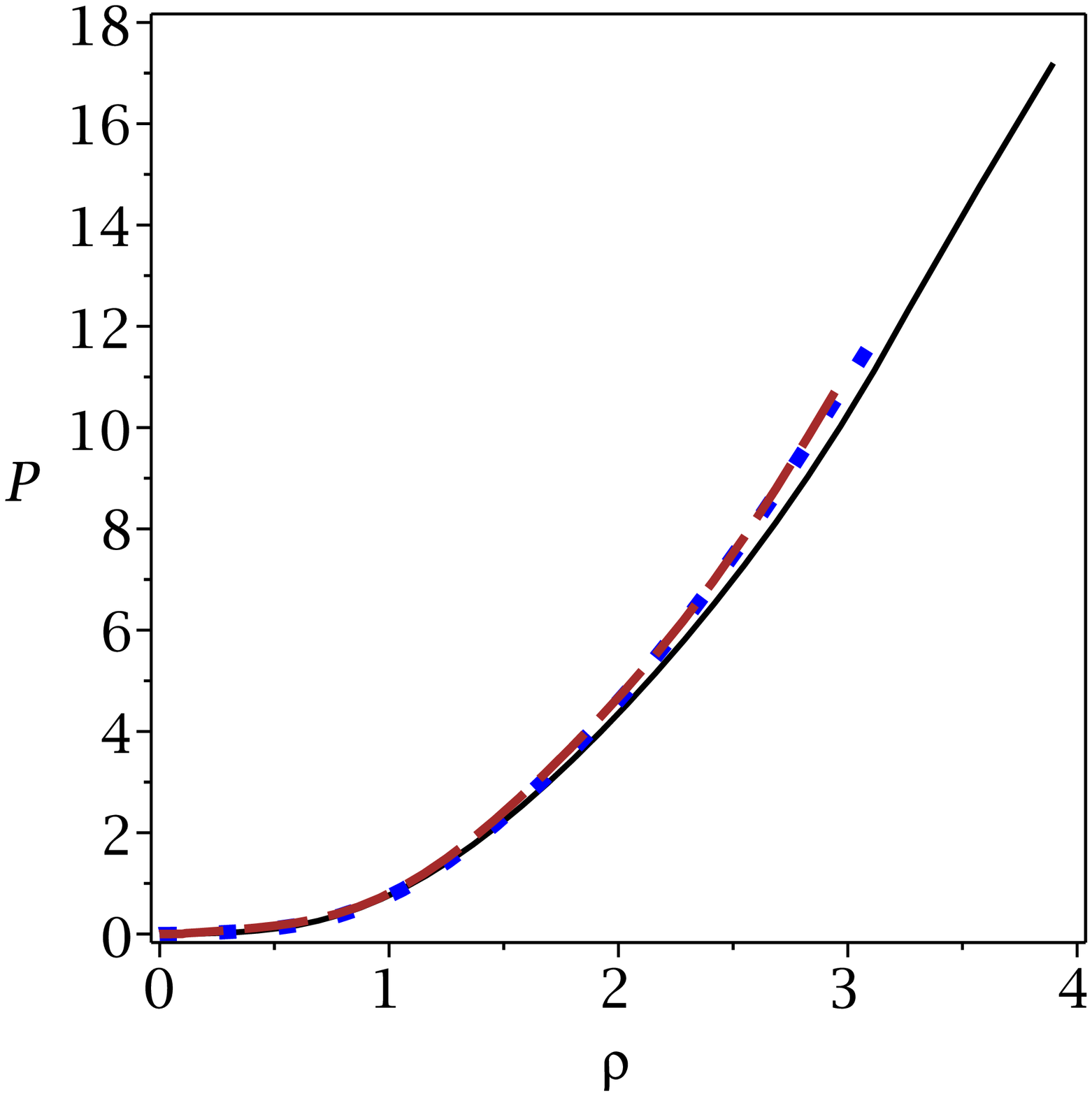} & \epsfxsize=8cm %
\epsffile{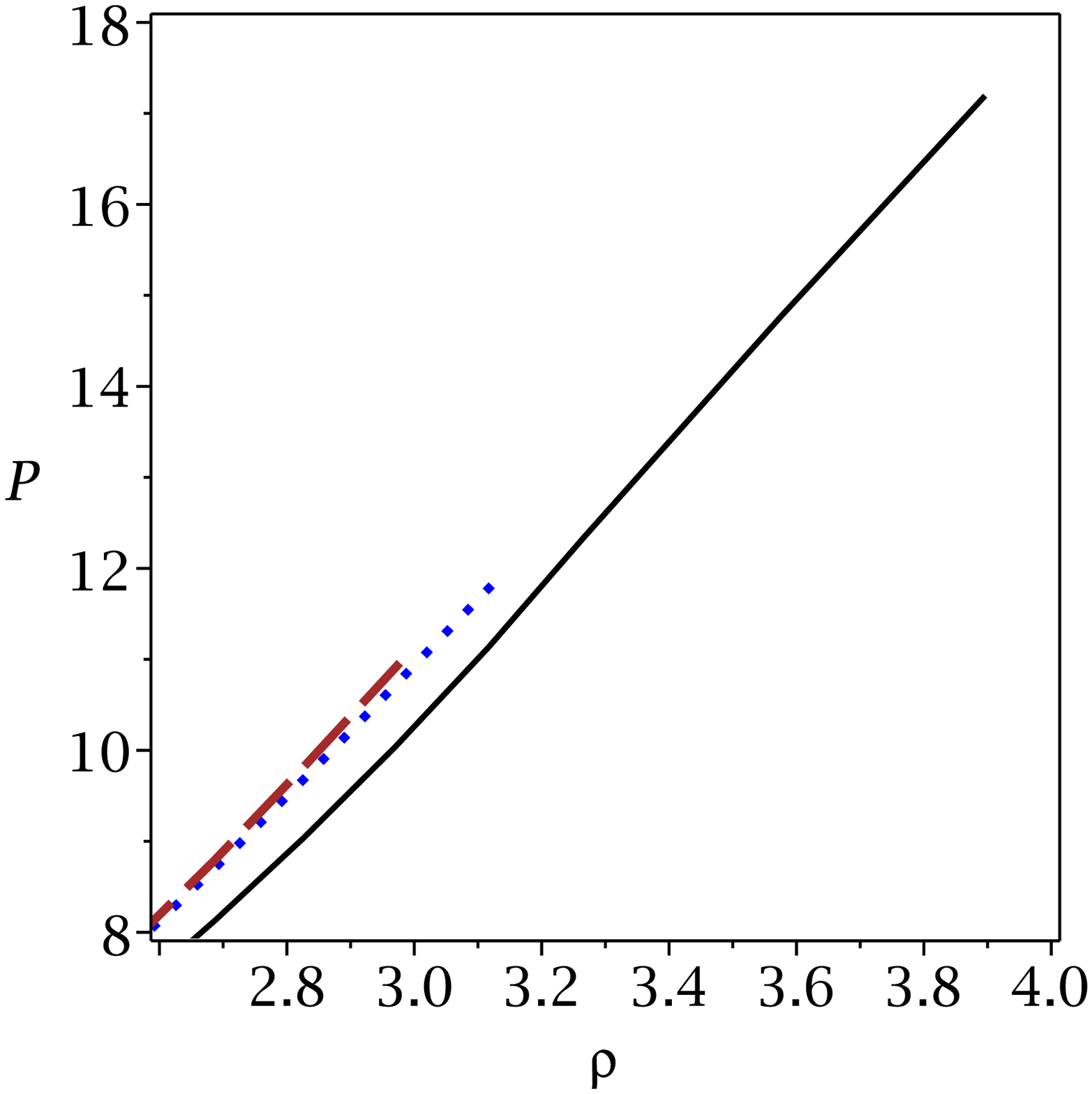}%
\end{array}
$%
\caption{EoS of magnetized neutron star matter (pressure, $P\times 10^{15}$
(g/$cm^{3}$) versus density ($\protect\rho \times 10^{15}$ (g/$cm^{3}$)) for
$B=0$ (continuous line), $B=5\times 10^{18}G$ (dotted line), and $B=1\times
10^{19}G$ (dashed line).}
\label{Fig1a}
\end{figure*}

From the energy of magnetized neutron matter, we have to consider a suitable
EoS. Here we evaluate the pressure using the following relation
\begin{equation*}
P(\rho ,B)=\rho ^{2}{\left( \frac{\partial E(\rho ,B)}{\partial \rho }%
\right) _{B}}.
\end{equation*}%
Our results for the EoS of magnetized neutron matter for different values of
the magnetic field have been presented in Fig. \ref{Fig1a}. It confirms that
the EoS of magnetized neutron matter becomes stiffer at higher magnetic
fields which is due to the inclusion of neutron anomalous magnetic moments.
\begin{figure}[tbp]
\epsfxsize=8cm \centerline{\epsffile{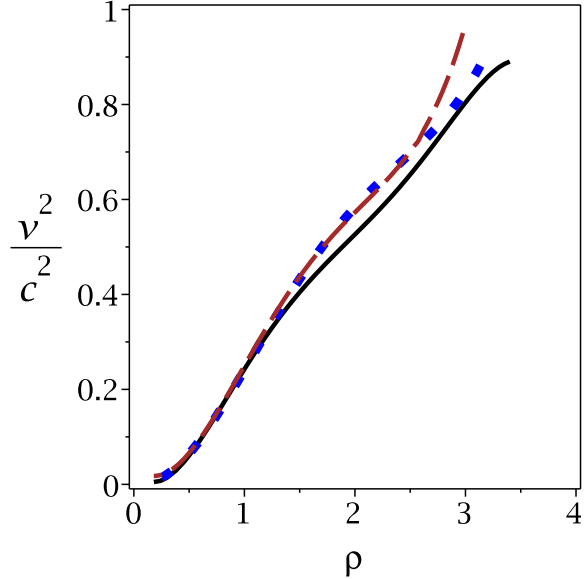}}
\caption{Sound speed ($v^{2}/c^{2}$) versus the density ($\protect\rho %
\times 10^{15}$ (g/$cm^{3}$)) for $B=0$ (continuous line), $B=5\times
10^{18}G$ (dotted line), and $B=1\times 10^{19}G$ (dashed line).}
\label{Fig2a}
\end{figure}

In order to do more investigations of the EoS introduced in this paper, we
analyze both energy and stability conditions. We examine the energy
conditions, such as null energy condition (NEC), weak energy condition
(WEC), strong energy condition (SEC), and also dominant energy condition
(DEC) at the center of neutron stars. The requirements of each energy
conditions can be summarize as
\begin{eqnarray}
NEC &\rightarrow &\ P_{c}+\rho _{c}\geq 0,  \label{11} \\
WEC &\rightarrow &\ P_{c}+\rho _{c}\geq 0,\ \ \ \&\ \ \rho _{c}\geq 0,
\label{22} \\
SEC &\rightarrow &\ P_{c}+\rho _{c}\geq 0,\ \ \ \&\ \ 3P_{c}+\rho _{c}\geq 0,
\label{33} \\
DEC &\rightarrow &\ \rho _{c}>\left\vert P_{c}\right\vert ,  \label{44}
\end{eqnarray}%
in which $\rho _{c}$ and $P_{c}$ are, respectively, the density and pressure
at the center ($r=0$)\ of a magnetized neutron star. Using Fig. \ref{Fig1a}
and the above conditions (\ref{11}-\ref{44}), we obtain the results related
to the energy conditions in table \ref{tab1}.
\begin{table*}[tbp]
\caption{Energy conditions for a neutron star.}
\label{tab1}
\begin{center}
\begin{tabular}{ccccccc}
\hline\hline
$B(G)$ & $\rho _{c}\ (10^{15}g/cm^3)$ & $P_{c}\ (10^{15}g/cm^3)$ & $NEC$ & $%
WEC$ & $SEC$ & $DEC$ \\ \hline\hline
$%
\begin{array}{c}
0.0 \\
5\times 10^{18} \\
1\times 10^{19}%
\end{array}
$ & $%
\begin{array}{c}
3.896 \\
3.134 \\
2.977%
\end{array}
$ & $%
\begin{array}{c}
1.910 \\
1.244 \\
1.218%
\end{array}
$ & $%
\begin{array}{c}
\checkmark \\
\checkmark \\
\checkmark%
\end{array}
$ & $%
\begin{array}{c}
\checkmark \\
\checkmark \\
\checkmark%
\end{array}
$ & $%
\begin{array}{c}
\checkmark \\
\checkmark \\
\checkmark%
\end{array}
$ & $%
\begin{array}{c}
\checkmark \\
\checkmark \\
\checkmark%
\end{array}
$ \\ \hline
&  &  &  &  &  &
\end{tabular}%
\end{center}
\end{table*}

Our results show that all energy conditions are satisfied (see table \ref%
{tab1}). So, this EoS of magnetized neutron star matter is suitable to study
the properties of magnetized neutron stars. In order to examine the
stability of EoS for a physically acceptable model, one expects that the
velocity of sound, $v$, be less than the light's velocity, $c$, \cite%
{Herrera,Abreu}. Considering the relation, $v^{2}=\left( \frac{dP}{d\rho }%
\right) $, we have presented the sound speed versus density in Fig. \ref%
{Fig2a}. It is evident that for all magnetic fields, the EoS of magnetized
neutron star matter satisfies the inequality $0\leq v^{2}\leq c^{2}$ (see
Fig. \ref{Fig2a}). Furthermore Le Chatelier's principle is another stability
condition that should be fulfilled. The establishment of this condition
indicates that both as a whole and also with respect to the non-equilibrium
elementary regions with spontaneous contraction or expansion, the object is
stable \cite{Ruffini,Glendenning}. It is evident from Fig. \ref{Fig1a} that
Le Chatelier's principle is satisfied. Therefore, the EoS of magnetized
neutron star matter satisfied both energy and stability conditions.





\begin{thebibliography}{99}
\bibitem{Ferrer} E. J. Ferrer, V. de la Incera, J. P. Keith, I. Portillo and
P. L. Springsteen, Phys. Rev. C \textbf{82}, 065802 (2010).

\bibitem{Yuan} Y. F. Yuan and J. L. Zhang, Astron. Astrophys. \textbf{335},
969 (1998).

\bibitem{Tatsumi} T. Tatsumi, Phys. Lett. B \textbf{489}, 280 (2000).

\bibitem{TolmanI} R. C. Tolman, Proc. Nat. Acad. Sc. \textbf{20}, 169 (1934).

\bibitem{TolmanII} R. C. Tolman, Phys. Rev. \textbf{55}, 364 (1939).

\bibitem{Oppenheimer} J. R. Oppenheimer and G. M. Volkoff, Phys. Rev.
\textbf{55}, 374 (1939).

\bibitem{Silbar} R. R. Silbar and S. Reddy, Am. J. Phys. \textbf{72}, 892
(2004).

\bibitem{Narain} G. Narain, J. Schaffner-Bielich and I. N. Mishustin, Phys.
Rev. D \textbf{74}, 063003 (2006).

\bibitem{BordbarBY} G. H. Bordbar, M. Bigdeli and T. Yazdizade, Int. J. Mod.
Phys. A \textbf{21}, 5991 (2006).

\bibitem{BoonsermVW} P. Boonserm, M. Visser and S. Weinfurtner, Phys. Rev. D
\textbf{76}, 044024 (2007).

\bibitem{LiX} X. Li, F. Wang and K. S. Cheng, JCAP \textbf{10}, 031 (2012).

\bibitem{Oliveira} A. M. Oliveira, H. E. S. Velten, J. C. Fabris and I. G.
Salako, Eur. Phys. J. C \textbf{74}, 3170 (2014).

\bibitem{He} X. T. He, F. J. Fattoyev, B. A. Li and W. G. Newton, Phys. Rev.
C \textbf{91}, 015810 (2015).

\bibitem{Yunes} N. Yunes and M. Visser, Int. J. Mod Phys. A \textbf{18},
3433 (2003).

\bibitem{Heyl} J. S. Heyl and L. Hernquist, Mon. Not. R. Astron. Soc.
\textbf{324}, 292 (2001).

\bibitem{Cardall} C. Y. Cardall, M. Prakash and J. M. Lattimer, Astrophys.
J. \textbf{554}, 322 (2001).

\bibitem{Kiuchi} K. Kiuchi, M. Shibata and S. Yoshida, Phys. Rev. D \textbf{%
78}, 024029 (2008).

\bibitem{Ciolfi} R. Ciolfi, V. Ferrari and L. Gualtieri, Mon. Not. R.
Astron. Soc. \textbf{406}, 2540 (2010).

\bibitem{Lopes} L. L. Lopes and D. P. Menezes, Braz. J. Phys. \textbf{42},
428 (2012).

\bibitem{Chirenti} C. Chirenti and J. Skakala, Phys. Rev. D \textbf{88},
104018 (2013).

\bibitem{Cheoun} M. K. Cheoun, C. Deliduman, C. G\"{u}ng\"{o}r, V. Keles, C.
Y. Ryu, T. Kajino and G. J. Mathews, JCAP \textbf{10}, 021 (2013).

\bibitem{Belvedere} R. Belvedere, J. A. Rueda and R. Ruffini, Astrophys. J.
\textbf{799}, 23 (2015).

\bibitem{Neilsen} D. Neilsen, S. L. Liebling, M. Anderson, L. Lehner, E.
O'Connor and C. Palenzuela, Phys. Rev. D \textbf{89}, 104029 (2014).

\bibitem{LopesDM} L. L. Lopes and D. P. Menezes, JCAP \textbf{08}, 002
(2015).

\bibitem{Kamiab} F. Kamiab, A. E. Broderick and N. Afshordi, arXiv:1503.03898

\bibitem{Broderick} A. E. Broderick, M. Prakash and J. M. Lattimer,
Astrophys. J. \textbf{537}, 351 (2000).

\bibitem{Astashenok} A. V. Astashenok, S. Capozziello and S. D. Odinstov,
Astrophys. Space Sci. \textbf{355}, 333 (2015).

\bibitem{PerlmutterI} S. Perlmutter et al., Astrophys. J. \textbf{517}, 565
(1999).

\bibitem{PerlmutterII} S. Perlmutter, M. S. Turner and M. White, Phys. Rev.
Lett. \textbf{83}, 670 (1999).

\bibitem{Riess} A. G. Riess et al., Astrophys. J. \textbf{607}, 665 (2004).

\bibitem{Horava} P. Horava, Phys. Rev. Lett. \textbf{102}, 161301 (2009).

\bibitem{MagueijoII} J. Magueijo and L. Smolin, Class. Quantum Grav. \textbf{%
21}, 1725 (2004).

\bibitem{GarattiniSar} R. Garattini and E. N. Saridakis, Eur. Phys. J. C
\textbf{75}, 343 (2015).

\bibitem{Amelino-Camelia} G. Amelino-Camelia, Nature. \textbf{418}, 34
(2002).

\bibitem{Peng} J. J. Peng and S. Q. Wu, Gen. Relativ. Gravit. \textbf{40},
2619 (2008).

\bibitem{Ali} A. F. Ali, Phys. Rev. D \textbf{89}, 104040 (2014).

\bibitem{HendiFEP} S. H. Hendi, M. Faizal, B. Eslam Panah and S. Panahiyan,
Eur. Phys. J. C \textbf{76}, 296 (2016).

\bibitem{HendiPEM} S. H. Hendi, S. Panahiyan, B. Eslam Panah and M.
Momennia, Eur. Phys. J. C \textbf{76}, 150 (2016).

\bibitem{LingLZ} Y. Ling, X. Li and H. Zhang, Mod. Phys. Lett. A \textbf{22}%
, 2749 (2007).

\bibitem{LiH} H. Li, Y. Ling and X. Han, Class. Quantum Grav. \textbf{26},
065004 (2009).

\bibitem{AliFM} A. F. Ali, M. Faizal and B. Majumder, EPL \textbf{109},
20001 (2015).

\bibitem{Galan} P. Galan and G. A. Mena Marugan, Phys. Rev. D \textbf{74},
044035 (2006).

\bibitem{Gim} Y. Gim and W. Kim, JCAP \textbf{05}, 002 (2015).

\bibitem{NonsingularI} A. Awad, A. F. Ali and B. Majumder, JCAP \textbf{10},
052 (2013).

\bibitem{NonsingularII} S. H. Hendi, M. Momennia, B. Eslam Panah and M.
Faizal, Astrophys. J. \textbf{827},153 (2016).

\bibitem{NonsingularIII} Y. Ling, JCAP \textbf{08}, 017 (2007).

\bibitem{NonsingularIV} S. H. Hendi, M. Momennia, B. Eslam Panah and S.
Panahiyan, Phys. Dark Universe \textbf{16}, 26 (2017).

\bibitem{Shapiro} S. Shapiro and S. Teukolsky, \textit{Black Holes, White
Dwarfs and Neutron Stars}. Wiley, New York (1983).

\bibitem{Psaltis} D. Psaltis, Living Rev. Relativ. \textbf{11}, 9 (2008).

\bibitem{Eksi} K. Y. Eksi, C. Gungor and M. M. Turkoglu, Phys. Rev. D
\textbf{89}, 063003 (2014).

\bibitem{Chandrasekhar} S. Chandrasekhar, Astrophys. J. \textbf{140}, 417
(1964).

\bibitem{BardeenTM} J. M. Bardeen, K. S. Thonre and D. W. Meltzer,
Astrophys. J. \textbf{145}, 505 (1966).

\bibitem{Kuntsem} H. Kuntsem, Mon. Not. R. Astron. Soc. \textbf{232}, 163
(1988).

\bibitem{Mak} M. K. Mak and T. Harko, Eur. Phys. J. C \textbf{73}, 2585
(2013).

\bibitem{Kalam} M. Kalam, S. M. Hossein and S. Molla, arXiv:1510.07015

\bibitem{Buchdahl} H. A. Buchdahl, Phys. Rev. \textbf{116}, 1027 (1959).

\bibitem{Moradi:2017alp} R. Moradi,~C. Stahl,~J. Firouzjaee and~S. S. Xue,
arXiv:1705.04168

\bibitem{Ray:2003gt} S. Ray,~A. L. Espindola,~M. Malheiro,~J. P. S. Lemos
and~V. T. Zanchin, Phys. Rev. D \textbf{68}, 084004 (2003).

\bibitem{HendiBEP} S. H. Hendi, G. H. Bordbar, B. Eslam Panah and S.
Panahiyan, JCAP \textbf{09}, 013 (2016).

\bibitem{AmelinoNature} G. Amelino-Camelia, J. R. Ellis, N. Mavromatos, D.
V. Nanopoulos and S. Sarkar, Nature. \textbf{393}, 763 (1998).

\bibitem{Jacob} U. Jacob, F. Mercati, G. Amelino-Camelia and T. Piran, Phys.
Rev. D \textbf{82}, 084021 (2010).

\bibitem{AmelinoLRR} G. Amelino-Camelia, Living Rev. Relativ. \textbf{16}, 5
(2013).

\bibitem{MagueijoI} J. Magueijo and L. Smolin, Phys. Rev. Lett. \textbf{88},
190403 (2002).

\bibitem{Rawlset} M. L. Rawls et al., Astrophys. J. \textbf{730}, 25 (2011).

\bibitem{Demorest} P. Demorest, T. Pennucci, S. Ransom, M. Roberts and J.
Hessels, Nature. \textbf{467}, 1081 (2010).

\bibitem{Antoniadis} J. Antoniadis et al. Science. \textbf{340}, 6131 (2013).

\bibitem{Clark et al} J. S. Clark et al., Astron. Astrophys. \textbf{392},
909 (2002).

\bibitem{Freire} P. C. C. Freire et al., Astrophys. J. \textbf{675}, 670
(2008).

\bibitem{Ruffini} Jr. C. E. Rhoades and R. Ruffini, Phys. Rev. Lett. \textbf{%
32}, 324 (1974).

\bibitem{HaeselII} P. Haensel, J. P. Lasota and J. L. Zdunik, Astron.
Astrophys. \textbf{344}, 151 (1999).

\bibitem{anisoI} C. G. Boehmer and T. Harko, Class. Quantum Grav. \textbf{23}%
, 6479 (2006).

\bibitem{anisoIII} B. C. Paul and R. Deb, Astrophys. Space Sci. \textbf{354}%
, 421 (2014).

\bibitem{anisoIV} S. K. Maurya, Y. K. Gupta, S. Ray and B. Dayanandan, Eur.
Phys. J. C \textbf{75}, 225 (2015).

\bibitem{anisoV} D. K. Matondo and S. D. Maharaj, Astrophys. Space Sci.
\textbf{361}, 221 (2016).

\bibitem{anisoVI} B. S. Ratanpal, V. O. Thomas and D. M. Pandya, Astrophys.
Space Sci. \textbf{361}, 65 (2016).

\bibitem{rotI} N. Andersson and G. L. Comer, Class. Quantum Grav. \textbf{18}%
, 969 (2001).

\bibitem{rotIII} Z. B. Etienne, Y. T. Liu and S. L. Shapiro, Phys. Rev. D
\textbf{74}, 044030 (2006).

\bibitem{rotVI} P. Pani and E. Berti, Phys. Rev. D \textbf{90}, 024025
(2014).

\bibitem{rotVII} R. F. P. Mendes, G. E. A. Matsas and D. A. T. Vanzella,
Phys. Rev. D \textbf{90}, 044053 (2014).

\bibitem{rotVIII} K. V. Staykov, D. D. Doneva, S. S. Yazadjiev and K. D.
Kokkotas, JCAP \textbf{10}, 006 (2014).

\bibitem{rotIX} A. Cisterna, T. Delsate, L. Ducobu and M. Rinaldi, Phys.
Rev. D \textbf{93}, 084046 (2016).

\bibitem{rotXI} J. G. Coelho, D. L. C\'{a}ceres, R. C. R. de Lima, M.
Malheiro, J. A. Rueda and R. Ruffini, Astron. Astrophys. \textbf{599}, A87
(2017).

\bibitem{rapidI} D. Lai and S. L. Shapiro, Astrophys. J. \textbf{442}, 259
(1995).

\bibitem{rapidII} S. Yoshida, S. Karino, S. Yoshida and Y. Eriguchi, Mon.
Not. R. Astron. Soc. \textbf{316}, L1 (2000).

\bibitem{rapidIII} D. D. Doneva, S. S. Yazadjiev, N. Stergioulas and K. D.
Kokkotas, Phys. Rev. D \textbf{88}, 084060 (2013).

\bibitem{rapidV} B. Kleihaus, J. Kunz, S. Mojica and M. Zagermann, Phys.
Rev. D \textbf{93}, 064077 (2016).

\bibitem{rapidVII} F. Cipolletta, C. Cherubini, S. Filippi, J. A. Rueda and
R. Ruffini, Phys. Rev. D \textbf{96}, 024046 (2017).

\bibitem{BordbarRM} G. H. Bordbar, Z. Rezaei and A. Montakhab, Phys. Rev. C
\textbf{83}, 044310 (2011).

\bibitem{Rezaei} Z. Rezaei and G. H. Bordbar, Eur. Phys. J. A \textbf{52},
132 (2016).

\bibitem{Herrera} L. Herrera, Phys. Lett. A \textbf{165}, 206 (1992).

\bibitem{Abreu} H. Abreu, H. Hernandez and L. A. Nunes, Class. Quantum Grav.
\textbf{24}, 4631 (2007).

\bibitem{Glendenning} N. K. Glendenning, Phys. Rev. Lett. \textbf{85}, 1150
(2000).
\end{thebibliography}
\end{document}